\documentclass[pss,fleqn,embeddedheads]{w-art}
\usepackage{times}
\usepackage{w-thm}
\usepackage{amssymb,amsmath,psfrag,url,color}

\newcommand{\MyField}[1]{{\bf{#1}}}
\newcommand{\MyCurl}{\nabla \times}

\newcommand{\MyTensor}[1]{{{{#1}}}}
\newcommand{\MySField}[1]{{{#1}}}
\newcommand{\MyMatrix}[1]{\mathrm{#1}}

\newcommand{\Problem}{{\bf Problem:}}
\newcommand{\epsInv}{\epsilon^{-1}}
\newcommand{\muInv}{\mu^{-1}}
\newcommand{\eps}{\epsilon}
\newcommand{\RotRotE}{\epsInv\MyCurl\muInv\MyCurl\MyField{E}}
\newcommand{\E}{\MyField{E}}
\newcommand{\MW}[1]{\epsInv\MyCurl\muInv\MyCurl\MyField{E}_{{#1}}-\omega^{2}\E_{{#1}}=0}
\newcommand{\MWStar}[1]{\nabla_{\xi\eta}\times{\mu_{*}}^{-1}\nabla_{\xi\eta}\times\MyField{E}_{{#1}}-\omega^{2}\eps_{*}\E_{{#1}}=0}
\usepackage[]{graphicx}
\begin{document}
%%    The information for the title page will be placed between
%%    \begin{document} and \maketitle. The order of most entries
%%    is determined by the class file and can not be changed by
%%    rearranging them. The maketitle command follows after the
%%    absract.
%%
%%    Most of the following commands will be completed by the publisher.
%%
\DOIsuffix{theDOIsuffix}
%%
%% issueinfo for header and copyright line
\Volume{XX}
\Issue{1}
\Copyrightissue{01}
\Month{01}
\Year{2004}
%%
%%    First and last pagenumber of the article. If the option
%%    'autolastpage' is set (default) the second argument may be left empty.
\pagespan{1}{}
%%
%%    Dates will be filled in by the publisher. The 'reviseddate' and
%%    'dateposted' (Published online) entry may be left empty.
%\Receiveddate{\sf zzz} \Reviseddate{\sf zzz} \Accepteddate{\sfzzz} \Dateposted{\sf zzz}
%%
%%    Give a maximum of six PACS code in numerical order.

%% \pretitle{Feature Article}

%% We have a short and a long form for the title. The short form
%% (optional argument) goes into the running head.

\title[Adaptive FEM for simulation of optical nano structures]{Adaptive Finite Element Method for Simulation of Optical Nano Structures}

%% Please do not enter footnotes or \inst{}-notes into the optional
%% argument of the author command. The optional argument will go into
%% the header.  If there is only one address the marker \inst{x} may be
%% omitted.

%% Information for the first author.
\author[J. Pomplun]{Jan Pomplun\footnote{Corresponding
     author: e-mail: {\sf pomplun@zib.de}, Phone: +49\,30\,841\,85\,273}}
\address{Zuse Institute Berlin,
Takustra{\ss}e 7,
14\,195 Berlin,
Germany}
\author[S. Burger]{Sven Burger}
\author[L. Zschiedrich]{Lin Zschiedrich}
\author[F. Schmidt]{Frank Schmidt}

\begin{abstract}
We discuss realization, properties and performance of the adaptive finite element approach to the design of nano-photonic components. Central issues are the construction of vectorial finite elements and the embedding of bounded components into the unbounded and possibly heterogeneous exterior. We apply the finite element method to the optimization of the design of a hollow core photonic crystal fiber. Thereby we look at the convergence of the method and discuss automatic and adaptive grid refinement and the performance of higher order elements.
\end{abstract}
%% maketitle must follow the abstract.
\maketitle                   % Produces the title.
\noindent
Published in phys. stat. sol. (b) \textbf{244}, No. 10, 3419-3434 (2007). \\\noindent
URL: http://www.interscience.wiley.com/
\section{Introduction}
The growing complexity and miniaturization of nano-optical components makes extensive simulations indispensable. Since in modern devices and applications the wavelength of light is of the same order as the dimension of the simulated structures Maxwell's equations have to be solved rigorously in order to get accurate results for the electromagnetic field. Examples of such structures are meta materials, photonic crystal devices, photolithographic masks and nano-resonators \cite{Burger2006b,HOL06,LindenEDKZKSBSW06,POM07,ZSC07}. A lot of different simulation techniques have been applied to and developed for nano-optical simulation, e.g. the finite element method (FEM), finite difference time domain simulations (FDTD), wavelet methods, finite integration technique (FIT), rigorously coupled wave analysis (RCWA), plane wave expansion methods (PWE).

Here we present the finite element method for the solution of time harmonic Maxwell's equations, i.e. we compute steady state solutions for electromagnetic fields with a fixed frequency $\omega$. 

This article is structured as follows. First we look at different problem classes which correspond to typical nano-optical simulation tasks and give the corresponding mathematical formulations of Maxwell's equations. These are propagation mode problems (Sec. \ref{sec:Propagation}), resonance problems (Sec. \ref{sec:Resonance}) and scattering problems (Sec. \ref{sec:Scattering}). In Section \ref{sec:Weak} we outline the weak formulation of Maxwell's equations which is needed for the discretization of Maxwell's equations with the finite element method. The basic ideas of the discretization are given in Sec. \ref{sec:discretization}. In the discretized version the solution to Maxwell's equations is determined in a subspace of the function space which contains the continuous solution. The construction of this subspace and therewith vectorial finite elements is given in Sec. \ref{sec:Contruction}. Since the finite size of nano-optical devices often has to be taken into account one needs to apply transparent boundary conditions to the computational domain. Section \ref{sec:Transparent} explaines our approach. It is based on the implementation of the perfectly matched layer (PML) method \cite{BerPML} and allows a certain class of inhomogeneous exterior domains \cite{Zschiedrich03}. The discretization scheme of the exterior domain is formulated in the context of the pole condition \cite{Schmidt02H}, which generalizes radiation conditions for wave propagation. In the final Section \ref{sec:Application} we apply the finite element method to the computation of leaky modes in hollow core photonic crystal fibers. We optimize the fiber design in order to minimize radiation losses.
\section{Problem Classes}
Many problems of light propagation can be formulated with time-harmonic Maxwell's equations. They can be divided into the following classes:
\begin{itemize}
\item Scattering problems: light scattering and transmission through arbitrary obstacles, e.g. meta materials, photo masks
\item Resonance problems: eigenmodes in resonators, e.g. cylindrical cavities, vertical cavity surface emitting lasers (VCSEL)
\item Propagation mode problems: guided light fields in waveguide structures, e.g. ridge waveguides, photonic crystal fibers
\end{itemize}
The basic equations for all of these classes are Maxwell's eigenvalue equations, which can be formulated as a second order curl curl equation for the electric field:
\begin{eqnarray}
  \label{eq:curlcurlE}
  \MW{}.
\end{eqnarray}
The finite element method is used to discretize this differential operator. Corresponding to the above classes different mathematical problem formulations arise due to different boundary conditions and unknown quantities which we will give in the following 3 sections. Resonance and propagating mode problems are eigenvalue problems where eigenmodes of the electric field and corresponding resonance frequencies respectively propagating constants have to be determined. In scattering problems an incident field of fixed frequency is given and the scattered light field from an arbitrarily shaped object has to be determined. 

The assembling of the finite element system however is very similar for all problem classes since always the same operator (\ref{eq:curlcurlE}) is discretized. After discretizing a scattering problem one has to solve a linear system of equations with the assembled finite element matrix and right hand side. After discretizing a propagating mode or resonance problem one has to solve an eigenvalue problem for the assembled finite element matrix.

%\section{Maxwell's equations using differential forms}
%In this section we want to use differential forms to derive the form of Maxwell's equations we are using for the finite element method. The time-harmonic source free Maxwell's equations read
%\begin{eqnarray}
%  \label{eq:mwAll}
%  \MyCurl\MyField{H}&=&i\omega\MyField{D}\\
%\MyCurl\E&=&-i\omega\MyField{B}\\
%\MyDiv\MyField{B}&=&0\\
%\MyDiv\MyField{D}&=&0
%\end{eqnarray}
%where $\E$ and $\MyField{H}$ are the electric and magnetic field $\MyField{B}$ is the magnetic inductance and $\MyField{D}$ the dielectric displacement. Furthermore we have the constitutive equations
%\begin{eqnarray}
%  \label{eq:mwConstl}
%\MyField{B}&=&\mu\MyField{H}\\
%\MyField{D}&=&\eps\E.
%\end{eqnarray}
%Now we use differential forms to state Maxwell's equations. They read
%\begin{eqnarray}
%  \label{eq:mwAllDiff}
%  d\MyField{H}&=&i\omega\MyField{D}\\
%d\E&=&-i\omega\MyField{B}\\
%d\MyField{B}&=&0\\
%d\MyField{D}&=&0
%\end{eqnarray}
%and the constitutive equations
%\begin{eqnarray}
%  \label{eq:mwConstlDiff}
%\MyField{B}&=&\star_{\mu}\MyField{H}\\
%\MyField{D}&=&\star_{\eps}\E,
%\end{eqnarray}
%where $d$ is the exterior derivative

\section{Propagation mode problems}
\label{sec:Propagation}
\begin{figure}[t]
\psfrag{x}{$x$}
\psfrag{y}{$y$}
\psfrag{z}{$z$}
\psfrag{omega}{$\Omega$}
\psfrag{gamma}{{\color{red} $\Gamma$}}
(a)\hspace{9.5cm}(b)\\
\includegraphics[height=3.2cm]{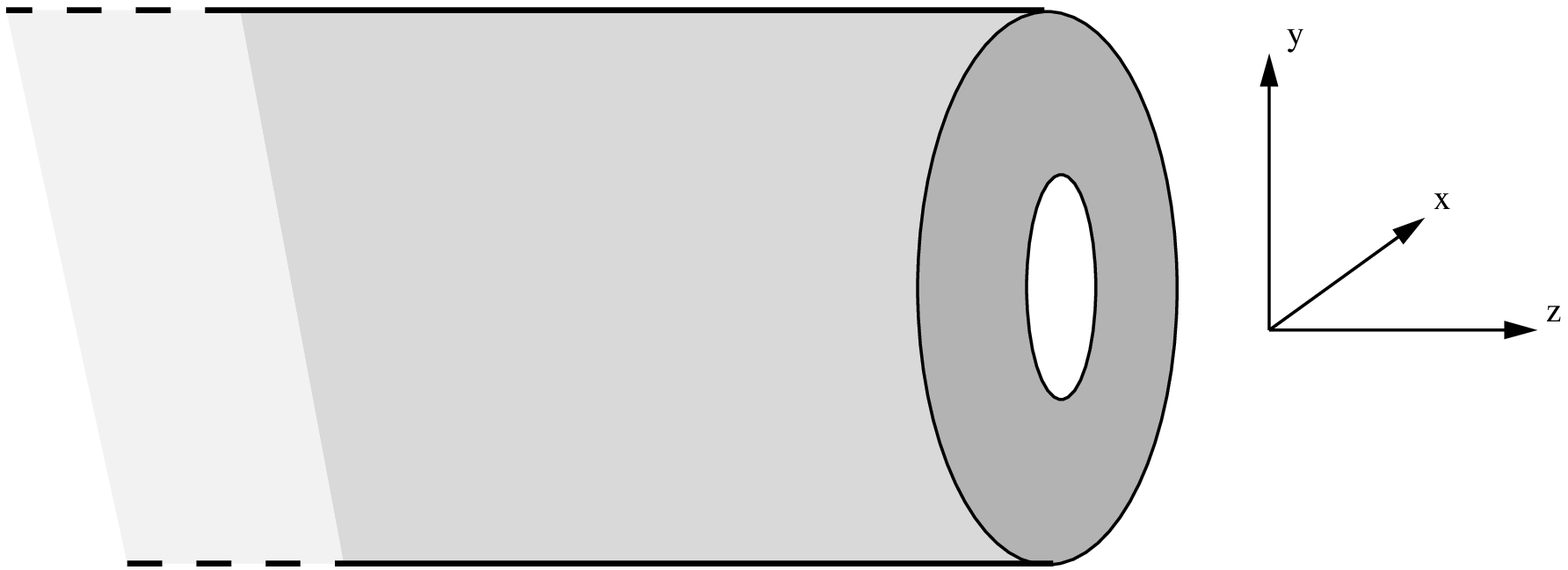}\hfill
\includegraphics[height=3.2cm]{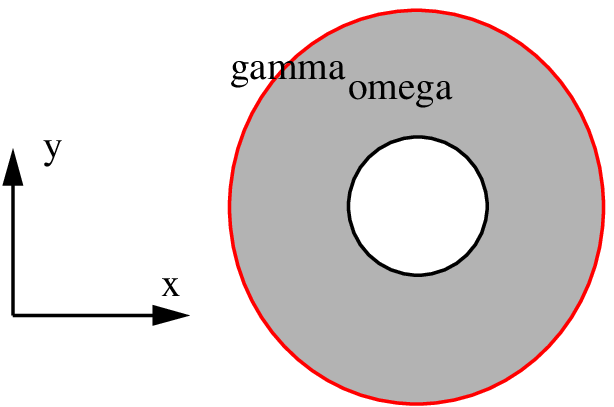}\hfill
\caption{\label{fig:fiberScheme}(a) 3D layout of a propagation mode problem. The waveguide structure has an invariance in one spatial dimension (we choose the $z$-direction). (b) 2D cross section of the waveguide structure which is sufficient for computation of propagating modes.}
\end{figure}
The geometry of a waveguide system $\Omega$ is invariant in one spatial dimension along the fiber, see Fig. \ref{fig:fiberScheme}(a). Here we choose the $z$-direction. Then a propagating mode is a solution to the time harmonic Maxwell's equations with frequency $\omega$, which exhibits a harmonic dependency in $z$-direction:
\begin{eqnarray}
\MyField{E}(x, y, z) & = & \MyField{E}_{\mathrm{pm}}(x, y)\exp \left(ik_{z}z\right)\nonumber\\
\MyField{H}(x, y, z) & = & \MyField{H}_{\mathrm{pm}}(x, y)\exp \left(ik_{z}z\right)\label{eq:propAnsatz}.
\end{eqnarray}
$\MyField{E}_{\mathrm{pm}}(x, y)$ and $\MyField{H}_{\mathrm{pm}}(x, y)$ are the electric and magnetic propagation modes and the parameter $k_{z}$ is called propagation constant. If the permittivity $\MyTensor{\epsilon}$ and permeability $\MyTensor{\mu}$ can be written as:
\begin{eqnarray}
\MyTensor{\epsilon}& =&
\left[
\begin{array}{cc}
\MyTensor{\epsilon}_{\perp\,\perp} & 0\\
0 & \MyTensor{\epsilon}_{zz}
\end{array}
\right]
\quad \mbox{and}\\
\MyTensor{\mu}& =&
\left[
\begin{array}{cc}
\MyTensor{\mu}_{\perp\,\perp} & 0 \\
0 & \MyTensor{\mu}_{zz} \label{eq:permitPermea}
\end{array}
\right],
\end{eqnarray}
we can split the propagation mode into a transversal and a longitudinal component:
\begin{equation}
\MyField{E}_{\mathrm{pm}}(x, y) =
\left[
\begin{array}{c}\MyField{E}_{\perp}(x, y) \\ \MySField{E}_{z}(x, y) \end{array}
\right].\label{eq:hprop}
\end{equation}
Inserting (\ref{eq:propAnsatz}) with (\ref{eq:permitPermea}) and (\ref{eq:hprop}) into Maxwell's equations yields:
\begin{equation}
\left[
\begin{array}{cc}
\MyMatrix{P} \nabla_{\perp} \MyTensor{\mu}_{zz}^{-1} \nabla_{\perp} \cdot \MyMatrix{P}
-k_z^2 \MyMatrix{P} \MyTensor{\mu}_{\perp\,\perp}^{-1} \MyMatrix{P}\, &
-ik_z \MyMatrix{P} \MyTensor{\mu}_{\perp\,\perp}^{-1} \MyMatrix{P} \nabla_{\perp} \\
-ik_z\nabla_{\perp}\cdot \MyMatrix{P} \MyTensor{\mu}_{\perp\,\perp}^{-1} \MyMatrix{P} &
\nabla_{\perp}\cdot \MyMatrix{P} \MyTensor{\mu}_{\perp\,\perp}^{-1} \MyMatrix{P} \nabla_{\perp}
\end{array}
\right]
\left[
\begin{array}{c}\MyField{E}_{\perp} \\ \MySField{E}_{z} \end{array}
\right] =
\left[
\begin{array}{cc}
\omega^{2}\MyTensor{\epsilon}_{\perp\,\perp} & 0 \\
0 & \omega^{2}\MyTensor{\epsilon}_{zz}
\end{array}
\right]
\left[
\begin{array}{c}\MyField{E}_{\perp} \\ \MySField{E}_{z} \end{array}
\right],\nonumber
\end{equation}
with
\begin{equation}
\MyMatrix{P} =
\left[
\begin{array}{cc}
0 & -1 \\
1 & 0
\end{array}
\right]
, \quad
\nabla_{\perp} =
\left[
\begin{array}{c}
\partial_x \\
\partial_y
\end{array}
\right].
\end{equation}
Now we define $\tilde{\MySField{E}}_{z}=k_{z}\MySField{E}_{z}$ and get the propagation mode problem:
\vspace{0.6cm}\\
\fbox{
\begin{minipage}[t]{\textwidth}
\Problem\\\noindent
Find tuples $(k_{z},\MyField{E}_{\perp},\MySField{E}_{z})$ such that:
\begin{eqnarray}
\MyMatrix{ A}
\left[
\begin{array}{c}
\MyField{E}_{\perp} \\
\tilde{\MySField{E}}_{z}
\end{array}
\right]
&=&
k_z^2 \,
\MyMatrix{ B}
\left[
\begin{array}{c}
\MyField{E}_{\perp} \\
\tilde{\MySField{E}}_{z}
\end{array}
\right]
\quad \mbox{in}\; {\mathbb R}^{2}.  \label{eq:evp}
\end{eqnarray}
with
\begin{eqnarray}
  \label{eq:ABMatrices}
  \MyMatrix{A}&=&\left[
\begin{array}{cc}
\MyMatrix{P} \nabla_{\perp} \MyTensor{\mu}_{zz}^{-1} \nabla_{\perp} \cdot \MyMatrix{P}
-\omega^{2}\MyTensor{\epsilon}_{\perp\,\perp}\, &
-i \MyMatrix{P} \MyTensor{\mu}_{\perp\,\perp}^{-1} \MyMatrix{P} \nabla_{\perp} \\
0 &
\nabla_{\perp}\cdot \MyMatrix{P} \MyTensor{\mu}_{\perp\,\perp}^{-1} \MyMatrix{P} \nabla_{\perp}- \omega^{2}\MyTensor{\epsilon}_{zz}
\end{array}
\right],\\
\MyMatrix{B}&=&\left[
\begin{array}{cc}
\MyMatrix{P} \MyTensor{\mu}_{\perp\,\perp}^{-1} \MyMatrix{P}
 & 0 \\
i\nabla_{\perp}\cdot \MyMatrix{P} \MyTensor{\mu}_{\perp\,\perp}^{-1} \MyMatrix{P} &
0
\end{array}
\right].
\end{eqnarray}
\end{minipage}}
\vspace{0.6cm}\\
Eq. (\ref{eq:evp}) is a generalized eigenvalue problem for the propagation constant $k_{z}$ and propagation mode $\MyField{E}_{\mathrm{pm}}(x, y)$. We get a similar equation for the magnetic field $\MyField{H}_{\mathrm{pm}}(x, y)$ exchanging $\MyTensor{\epsilon}$ and $\MyTensor{\mu}$. For the numerical analysis of a propagation mode problem (in Sec. \ref{sec:Application}) we furthermore define the effective refractive index $n_{\mathrm{eff}}$ which we will also refer to as eigenvalue:
\begin{eqnarray}
\label{eq:neff}
n_{\mathrm{eff}}& =& \frac{k_z}{k_0} \qquad \mbox{with}\\
k_0 &=& \frac{2\pi}{\lambda_0}\nonumber,
\end{eqnarray}
where $\lambda_{0}$ is the vacuum wavelength of light.
Note that we stated the propagating mode problem (\ref{eq:evp}) on ${\mathbb R}^{2}$, Fig. \ref{fig:fiberScheme}(b). This means that we take the infinite exterior of the waveguide into account, which allows to compute bounded as well as leaky modes. Leaky modes thereby model radiation losses from the waveguide to the exterior. Using the finite element method we therefore have to state transparent boundary conditions on $\Gamma$, see Fig. \ref{fig:fiberScheme}(b). We realize these boundary conditions with the PML method which also allows inhomogeneous exterior domains. We will explain our implementation of the PML in Sec. \ref{sec:Transparent}.

\section{Resonance problems}
\label{sec:Resonance}
\begin{figure}[ht]
\psfrag{omega}{$\Omega$}
\psfrag{gamma}{{\color{red}$\Gamma=\partial\Omega$}}
\psfrag{normal}{$\MyField{n}$}
\psfrag{perfectly}{$\mathbb{R}^{n}\setminus\Omega$}
\psfrag{omega}{$\Omega$}
\psfrag{normal}{$\MyField{n}$}
\psfrag{S}{$S$}
\psfrag{ein}{$\MyField E_{in}$}
\psfrag{eout}{$\MyField E_{out}$}
\psfrag{edrin}{$\MyField E$}
\psfrag{Rn}{$\mathbb{R}^{n}\setminus\Omega$}
\psfrag{normal}{$\MyField{n}$}
\psfrag{muext}{$\mu_{ext}$}
\psfrag{epsext}{$\epsilon_{ext}$}
(a)\hspace{7cm}(b)\\
\includegraphics[height=5cm]{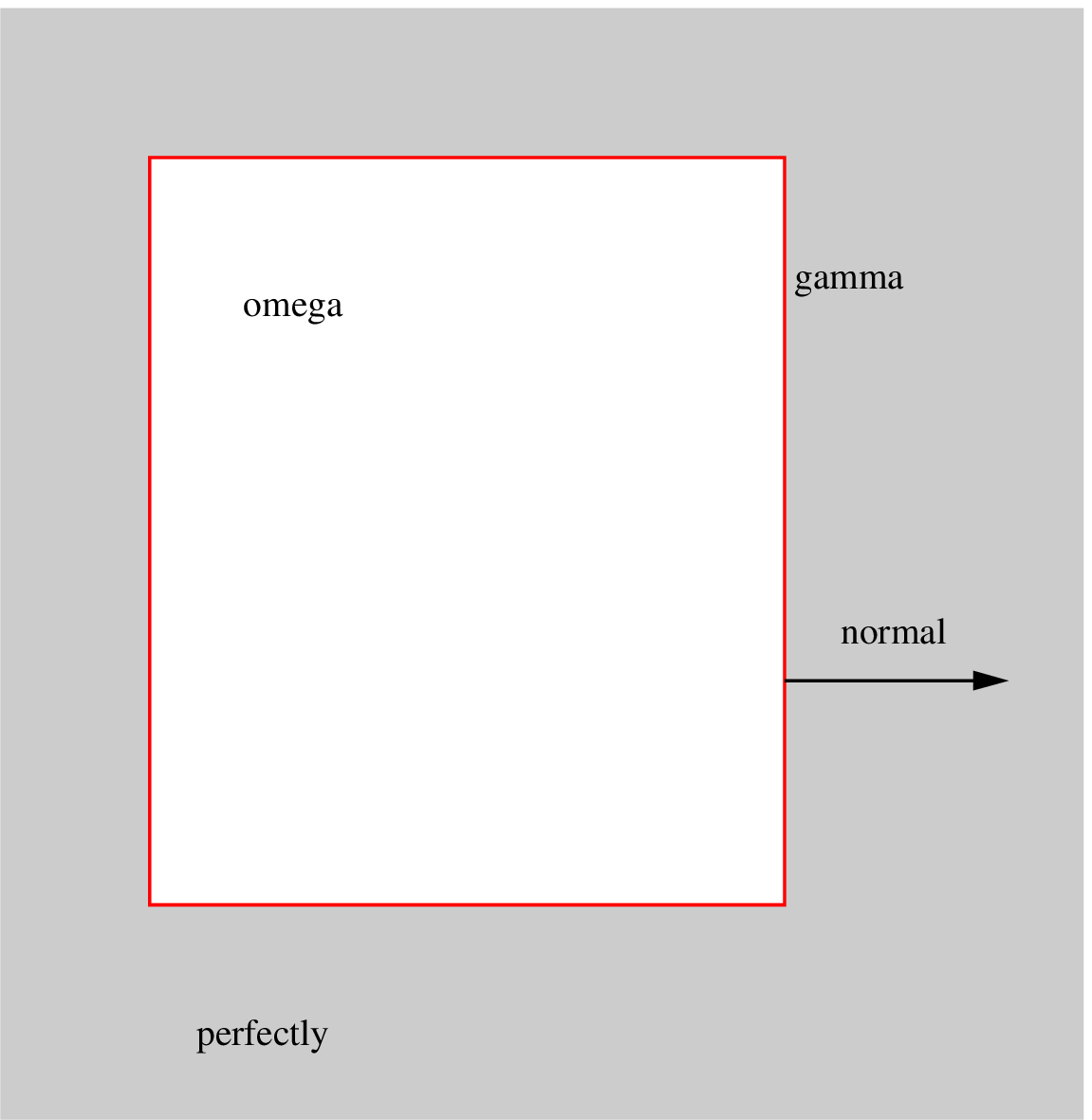}\hfill
\includegraphics[height=5cm]{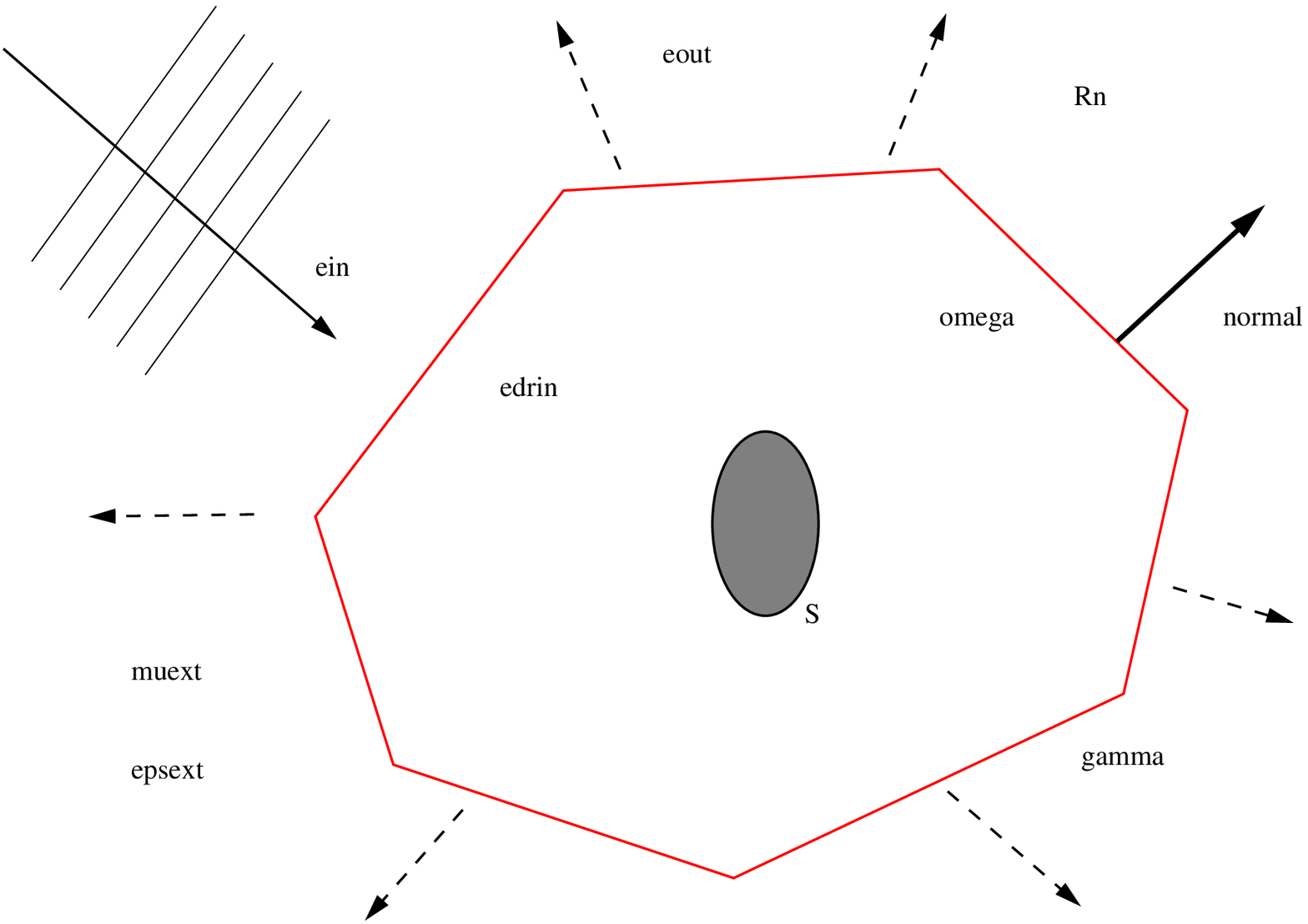}\hfill
\caption{\label{fig:resonance}(a) Setup for an electromagnetic resonance problem. The walls of the resonator are perfectly conducting. The components of the electric field perpendicular to the normal $\MyField{n}$ of the boundary $\Gamma$ therefore have to vanish. (b) Setup of a scattering problem. The interior domain $\Omega$ contains the scatterer $S$ and is embedded into an infinite exterior ${\mathbb{R}}^{n}$ with permittivity $\epsilon_{ext}$ and permeability $\mu_{ext}$. The incoming electric field $\MyField E_{in}$ is entering the interior domain via the boundary $\Gamma$ and is the source for the electric field inside $\Omega$. The scattered light $\MyField E_{out}$ is originated within $\Omega$. It is therefore strictly outward propagating.}
\end{figure}
Let us consider a cavity $\Omega$ with boundary $\Gamma=\partial \Omega$ as depicted in Fig. \ref{fig:resonance}(a). Inside $\Omega$ the electric field has to fulfill Maxwell's equations. On $\Gamma$ we have to impose boundary conditions. Since in nano-optical applications the coupling of a resonator to the exterior often can not be neglected we use transparent boundary conditions. \\
\noindent
Now we are interested in resonance modes and corresponding frequencies in the cavity:
\vspace{0.5cm}\\
\fbox{
\begin{minipage}[t]{\textwidth}
\Problem\\\noindent
Find tuples $(\omega,\MyField{E})$ such that:\\\noindent
Maxwell's equation in interior domain:
\begin{eqnarray}
  \label{eq:resonanceProblem}
  \RotRotE&=&\omega^{2}\E\;\;\mbox{in}\;\Omega
\end{eqnarray}
Maxwell's equation in exterior domain:
\begin{eqnarray}
  \label{eq:resonanceProblemOut}
  \MW{out}\;\;\mbox{in}\;{\mathbb R}^{n}\setminus\Omega
\end{eqnarray}
Boundary condition at $\Gamma$:
\begin{eqnarray}
  \label{eq:bcmw1res}
\left({\MyTensor{\mu}}^{-1}\MyCurl\MyField{E}_{{out}}\right)\times\MyField{n}&=&\left({\MyTensor{\mu}}^{-1}\MyCurl\MyField{E}\right)\times\MyField{n}\\
\MyField{E}_{{out}}\times\MyField{n}&=&\MyField{E}\times\MyField{n},
\end{eqnarray}
Silver-M\"uller radiation condition (``boundary condition at infinity'') for exterior field:
\begin{eqnarray}
  \label{eq:smmwres}
&&\lim\limits_{r\rightarrow\infty}r\left( \MyCurl\E_{out}(\MyField{r})\times\MyField{r}_{0}-i\frac{\omega\sqrt{\epsilon_{ext}\mu_{ext}}}{c}\MyCurl\E_{out}(\MyField{r})  \right)=0\\
&&\mbox{uniformly continuous in each direction $\MyField{r}_{0}$,}\nonumber
\end{eqnarray}
where $\MyField{r}$ the coordinate vector in ${\mathbb R}^{n}$, $r$ its norm and $\MyField{r}_{0}=\frac{\MyField{r}}{r}$.
\end{minipage}}
\vspace{0.5cm}\\
Since all fields $\E$ which are gradients of a scalar potential $\Phi$ lie in the kernel of the curl operator, i.e. $ \MyCurl \E=0\Rightarrow\E=\nabla\Phi$, the above problem has many solutions to $\omega=0$. For a numerical method it is important to guarantee that the approximated (discrete) fields which correspond to these gradient fields also have $\omega=0$. For the finite element method this means that one has to construct carefully appropriate ansatz function in order to preserve the mathematical structure of Maxwell's equations in the discrete version. We will come to this point in more detail in Sec. \ref{sec:discretization}.

\section{Scattering problems}
\label{sec:Scattering}
The setup of a scattering problem is depicted in Fig. \ref{fig:resonance}(b). The region in space occupied with the scatterer $S$ is denoted by $\Omega$. For simplicity we assumed here that we have a homogeneous exterior with relative permittivity $\epsilon_{ext}$ and permeability $\mu_{ext}$. In the exterior ${\mathbb R}^{n}$ we have an incident field $\MyField{ E}_{in}$ which enters the interior domain $\Omega$ across its boundary $\Gamma=\partial\Omega$ and is scattered. The scattered field $\MyField {E}_{out}$ originates inside $\Omega$ and is therefore strictly outgoing. The total field is then $\MyField{E}=\MyField {E}_{in}+\MyField{E}_{out}$. The scattering problem is formulated as follows:
\vspace{0.5cm}\\
\fbox{
\begin{minipage}[t]{\textwidth}
\Problem\\\noindent
Maxwell's equation in interior domain:
\begin{eqnarray}
  \label{eq:mw}
  \MW{}\;\;\mbox{in}\;\Omega
\end{eqnarray}
Maxwell's equation in exterior domain:
\begin{eqnarray}
  \label{eq:mw1}
\epsInv\MyCurl\muInv\MyCurl\MyField{E}_{{out}}-\omega^{2}\E_{{out}}&=&0\;\;\mbox{in}\;{\mathbb R}^{n}\setminus\Omega
\end{eqnarray}
Boundary condition at $\Gamma$:
\begin{eqnarray}
  \label{eq:bcmw1}
\left({\MyTensor{\mu}}^{-1}\MyCurl(\MyField{E}_{in}+\MyField{E}_{{out}})\right)\times\MyField{n}&=&\left({\MyTensor{\mu}}^{-1}\MyCurl\MyField{E}\right)\times\MyField{n}\\
(\MyField{E}_{{in}}+\MyField{E}_{{out}})\times\MyField{n}&=&\MyField{E}\times\MyField{n},
\end{eqnarray}
where the incoming field $\MyField{E}_{in}$ has to fulfill Maxwell's equations in a neighborhood of the boundary $\Gamma$.
Silver-M\"uller radiation condition (``boundary condition at infinity'') for exterior field:
\begin{eqnarray}
  \label{eq:smmw1}
&&\lim\limits_{r\rightarrow\infty}r\left( \MyCurl\E_{out}(\MyField{r})\times\MyField{r}_{0}-i\frac{\omega\sqrt{\epsilon_{ext}\mu_{ext}}}{c}\MyCurl\E_{out}(\MyField{r})  \right)=0\\
&&\mbox{uniformly continuous in each direction $\MyField{r}_{0}$,}\nonumber
\end{eqnarray}
where $\MyField{r}$ the coordinate vector in ${\mathbb R}^{n}$, $r$ its norm and $\MyField{r}_{0}=\frac{\MyField{r}}{r}$.
\end{minipage}}
\vspace{0.5cm}\\

The Silver M\"uller radiation condition guarantees that $\E_{out}$ is a strictly outward radiating solution. For inhomogeneous exterior domains a generalization of this radiation condition has to be stated. A deeper understanding of the meaning of outward radiating fields is made by the pole condition \cite{Schmidt02H} which characterizes these fields by the poles of their Laplace transforms. Using the FEM method $\Omega$ is taken as computational domain. On $\Gamma$ transparent boundary conditions have to be stated.

\section{Weak formulation of Maxwell's equations}
\label{sec:Weak}
For application of the finite element method we have to derive a weak formulation of Maxwell's equations. We multiply (\ref{eq:mw}) with a vector valued test function
$\MyField{\Phi} \in V=H(curl,\Omega)$ \cite{MON03} and integrate over the domain $\Omega$:
\begin{eqnarray}
  \label{eq:MWweak1}
  \int_{\Omega}\left\{\overline{\MyField{\Phi}}\cdot\left[\nabla\times{\MyTensor{\mu}}^{-1}\nabla\times\MyField{ E}\right]-\omega^{2}\MyTensor{\epsilon}\, \overline{\MyField{\Phi}}\cdot\MyField{ E}\right\}d^{3}r=0\,,\;\forall {\MyField{\Phi}}\in V,
\end{eqnarray}
where bar denotes complex conjugation. After a partial integration we arrive at the weak formulation of Maxwell's equations:\\
Find $\MyField{ E}\in V=H(curl,\Omega)$ such that
\begin{eqnarray}
  \label{eq:MWweak2}
&&\int_{\Omega}\left\{\overline{\left(\nabla\times\MyField{\Phi}\right)}\cdot\left({\MyTensor{\mu}}^{-1}\nabla\times\MyField{ E}\right)-\omega^{2}\MyTensor{\epsilon}\, \overline{\MyField{\Phi}}\cdot\MyField{ E}\right\}d^{3}r=\int_{\Gamma}\overline{\MyField{\Phi}}\cdot\MyField{ F} d^{2}r\,,\;\forall \MyField{\Phi}\in V,
\end{eqnarray}
with
\begin{eqnarray}
  \label{eq:DiriData}
  &&\left({\MyTensor{\mu}}^{-1}\MyCurl\MyField{E}\right)\times\MyField{n}=\MyField{F}
\quad\mbox{given on $\Gamma$ (Neumann boundary
condition).}
\end{eqnarray}
Hence one needs Neumann data on $\Gamma$ for the electric field in order to solve the weak problem. Therefore one has to construct an operator which maps the Dirichlet data of the electric field onto its Neumann values respecting the radiation condition. The construction of such a Dirichlet to Neumann operator will be explained in Sec. \ref{sec:Transparent}. In order to state the weak formulation we define the following bilinear functionals:
\begin{eqnarray}
  \label{eq:forms}
a(\MyField w,\MyField v)&=&\int_{\Omega}\overline{\left(\MyField\nabla\times\MyField w\right)}\cdot\left({\MyTensor{\mu}}^{-1}\MyField\nabla\times\MyField v\right)-\MyTensor{\omega^{2}\epsilon}\,\overline{\MyField w}\cdot\MyField v \,d^{3}r,\label{defa}\\
f(\MyField w)&=&\int_{\Gamma}\overline{\MyField w}\cdot\MyField F d^{2}r\label{defF}
\end{eqnarray}
The weak formulation of Maxwell's equations then reads:
\vspace{0.5cm}\\
\fbox{
\begin{minipage}[t]{\textwidth}
Find $\MyField{v}\in V=H(curl,\Omega)$ such that
\begin{eqnarray}
  \label{eq:mwWeak}
 && a(\MyField w,\MyField v)=f(\MyField w)\,,\;\forall \MyField{w}\in V.
\end{eqnarray}
\end{minipage}}
\vspace{0.5cm}\\
The above equation is an exact reformulation of Maxwell's equations.

\section{Discretization of Maxwell's equations}
\label{sec:discretization}
\begin{figure}[t]
(a)\hspace{8cm}(b)\hspace{5cm}\\
\includegraphics[height=5.1cm]{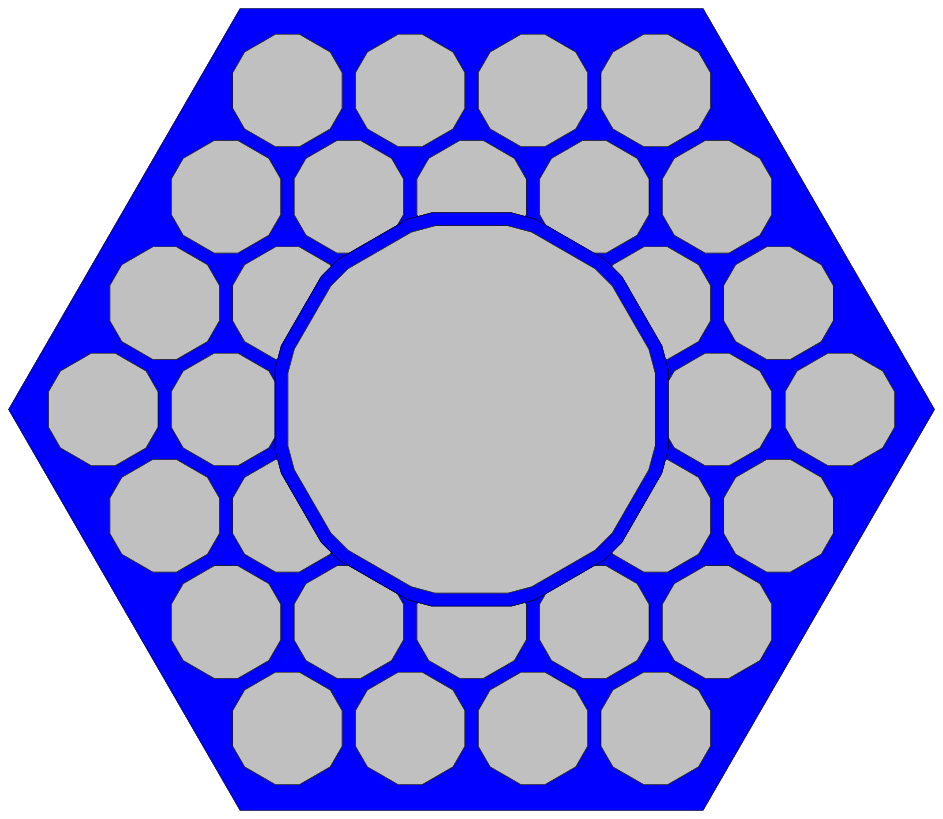}\hspace{0.3cm}\hfill
\includegraphics[height=5.1cm]{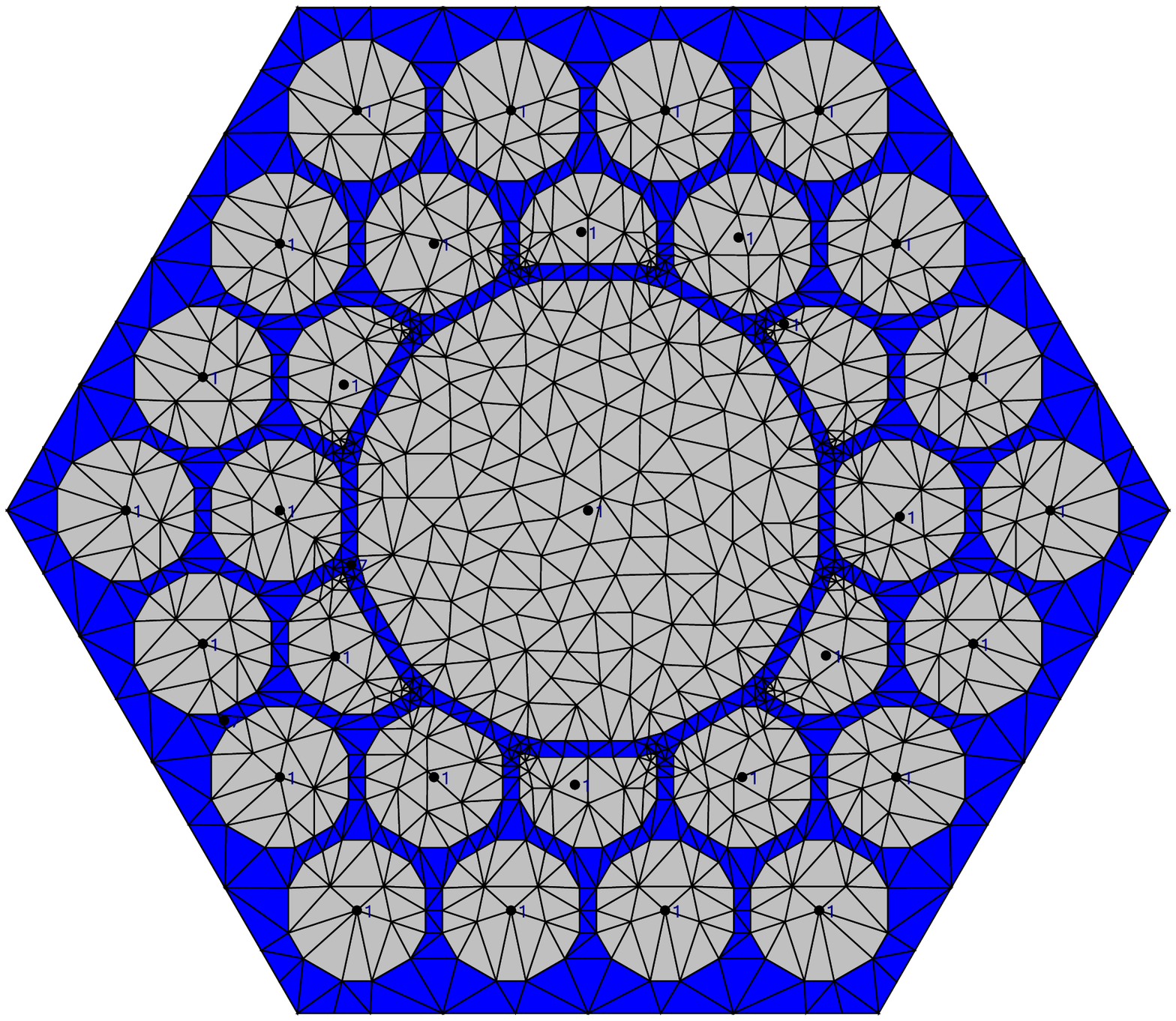}\hfill
\caption{\label{fig:fem}(a) Computational domain and (b) triangulation of photonic crystal fiber cross section.}
\end{figure}
Now we discretize the weak form of Maxwell's equations. The finite element method thereby restricts the space $V$ to a finite dimensional subspace $V_{h}\subset V$, $dim V_{h}=N_{h}$. This means the finite element method does not approximate Maxwell's equations itself but the function space in which the solution is determined. The finite dimensional finite element space is an approximation to the solution space of the continuous problem. The discretized Maxwell's version (compare to (\ref{eq:mwWeak})) simply reads:
\vspace{0.5cm}\\
\fbox{
\begin{minipage}[t]{\textwidth}
Find $\MyField{v}\in V_{h}\subset H(curl,\Omega)$ such that
\begin{eqnarray}
  \label{eq:mwWeakdisc}
 && a(\MyField w,\MyField v)=f(\MyField w)\,,\;\forall \MyField{w}\in V_{h}.
\end{eqnarray}
\end{minipage}}
\vspace{0.5cm}\\
 The subspace $V_{h}$ and therewith the basis for the approximate solution are constructed as follows. One starts with a computational domain $\Omega$ for example the cross section of a photonic crystal fiber (\ref{fig:fem})(a). This domain is subdivided into small patches, e.g. triangles or quadrilaterals in 2D and tetrahedrons in 3D, Fig \ref{fig:fem}(b). On these patches vectorial ansatz functions $\MyField{\varphi}_{i}$ are defined whose construction will be explained in section \ref{sec:Contruction}. These functions are usually polynomials of a fixed order whose support is restricted to one or a small number of patches. The set $\{\MyField{\varphi}_{1},\dots,\MyField{\varphi}_{N_{h}}\}$ of ansatz functions forms a basis of $V_{h}$. The approximate solution ${\MyField{E}_{h}}$ for the electric field is a superposition of these local ansatz functions:
\begin{eqnarray}
  \label{eq:EAnsatz}
  \MyField{E}_{h}=\sum_{i=1}^{N}a_{i}\MyField{\varphi}_{i}
\end{eqnarray}
If we insert this expansion of the electric field into the discrete version of Maxwell's equations (\ref{eq:mwWeakdisc}) for $\MyField v$ and replace $\forall \MyField{w}\in V_{h}$ by $\forall \MyField{w}\in \{ {\varphi}_{1},\dots,\varphi_{N_{h}}  \}$ (which is equivalent because this is a basis of $V_{h}$) the discrete Maxwell's equation reads:
\begin{eqnarray}
  \label{eq:MWweak3}
&&  \sum_{i=1}^{N}a_{i} a(\varphi_{j},\varphi_{i})=f(\varphi_{j})\,,\;\forall j=1,\dots,N
\end{eqnarray}
which is a linear system of equations for the unknown coefficients $a_{i}$:
\begin{eqnarray}
  \label{eq:MWweak4}
&&\MyMatrix{A}\cdot \vec a=\vec f
\end{eqnarray}
with $\MyMatrix A_{ji}=a(\varphi_{j},\varphi_{i}),\;f_{j}=f(\varphi_{j}),\;
\vec a=\left(
\begin{array}{c}
a_{1}\\
...\\
a_{N}
\end{array}
\right)$\\\noindent
The matrix entries $a(\varphi_{j},\varphi_{i})$ arise from computing integrals (\ref{defa}). Since the ansatz functions $\MyField{\varphi}_{i}$ have a small support the stiffness matrix $A_{ji}$ has $O(N)$ nonzeros out of $O(N^2)$ entries. The arising matrix is therefore sparse. Using special solvers the computational time scales practically linearly with the number of unknowns.
\section{Construction of finite elements}
\label{sec:Contruction}
In this section we want to show how appropriate ansatz functions $\varphi_{i}$ of the finite dimensional subspace $V_{h}$ are constructed for the finite element method. We restrict ourselves to the 2D case and consider triangles as patches which we will denote with $K$. The global function space $V_{h}$ with $dim V_{h}=N_{h}$ is separated into local function spaces $V_{K}$ with $dim V_{K}=N_{K}$ on each patch $K$. For the union of all patches we have $\cup K=\Omega$.

On each patch we now define $N_{K}$ basis functions $\varphi_{i}$, $i=1,\dots,N_{K}$. Furthermore we have to define $N_{K}$ functionals $\psi_{j}$ which are called degrees of freedom. These functionals are constructed such that
\begin{eqnarray}
  \label{eq:deg}
  \psi_{j}(\varphi_{i})=\delta_{ij}
\end{eqnarray}
is satisfied. The meaning of the degrees of freedom can be understood when considering the approximation of the solution on a patch, which is a superposition of the local functions $\varphi_{i}$:
\begin{eqnarray}
  \label{eq:approx}
&&  \sum\limits_{i=1}^{N_{K}}a_{i}\varphi_{i}\rightarrow\psi_{j}\left(\sum\limits_{i=1}^{N_{K}}a_{i}\varphi_{i}\right)=a_{j}.
\end{eqnarray}
Hence the degree of freedom $\psi_{j}$ returns the coefficient $a_{j}$ of the basis function $\varphi_{j}$.

Now we start constructing finite elements. Therefore we have to perform the following steps. Choose a patch $K$, e.g. triangle or quadrilateral. Define a function space $V_{K}$ on $K$ which has some desired properties. E.g. if the field which we are approximating is differentiable then the elements of $V_{K}$ should also be differentiable. Finally define the degrees of freedom $\psi_{j}$ and a set of basis functions $\varphi_{i}$ which span $V_{K}$.

We start with linear scalar finite elements of order $1$ on triangles. These elements are $H^{1}$ conform. A function in $H^{1}$ has a first weak derivative. Furthermore the function itself and its weak derivative are quadratic Lebesgue integrable \cite{MON03}. Let us consider a triangle with nodes $(x_{1},y_{1})$, $(x_{2},y_{2})$, $(x_{3},y_{3})$. The polynomial function space of order $1$ on this triangle is
\begin{eqnarray}
  \label{eq:h1Space}
  P^{1}(K)=\left\{v=a+bx+cy,\,a,b,c\in  {\mathbb R}        \right\}.
\end{eqnarray}
This is our local ansatz space with $dim P^{1}(K)=3$. Now we want to construct the basis functions in $P^{1}(K)$ which we will call $\lambda_{i}$ and degrees of freedom $\psi_{j}$. First we define the degrees of freedom. For a $v\in P^{1}(K)$ we choose:
\begin{eqnarray}
  \label{eq:dofP1}
  \psi_{i}(v)&:=&\int_{K}\delta\left[ (x,y)-(x_{i},y_{i})\right]vdxdy=v(x_{i},y_{i}),\; i=1,2,3\nonumber
\end{eqnarray}
hence the degrees of freedom simply give the value of a function of $P^{1}(K)$ at the nodes of the patch. Using $\psi_{j}(\lambda_{i})=\delta_{ij}$ (\ref{eq:deg}) we can construct a basis of $P^{1}(K)$.
\begin{figure}[t]
\psfrag{xl}{$x$}
\psfrag{yl}{$y$}
\psfrag{0}{$0$}
\psfrag{1}{$1$}
\psfrag{0.5}{$0.5$}
(a)\hspace{4.9cm}(b)\hspace{4.7cm}(c)\\
\includegraphics[height=3.9cm]{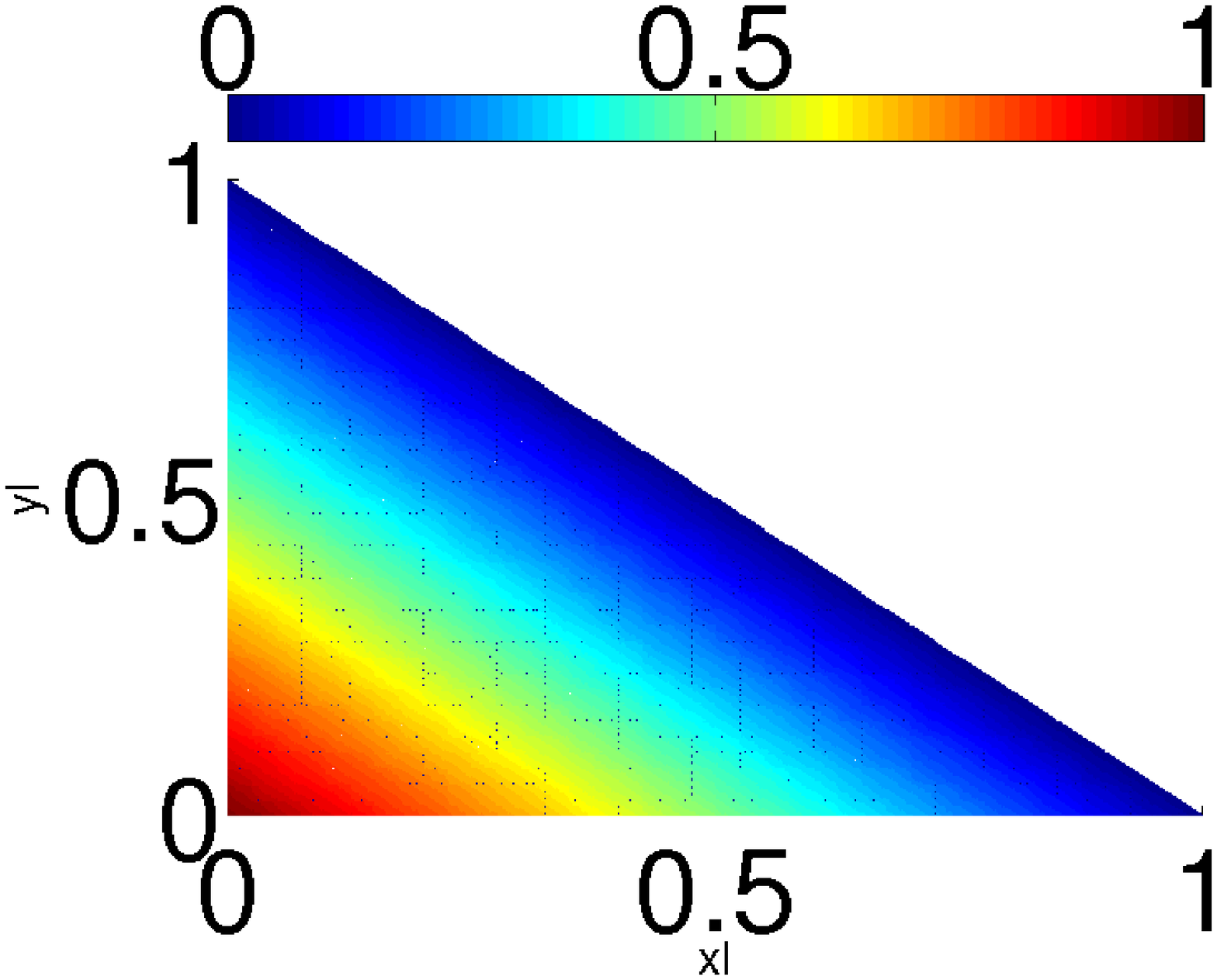}\hspace{0.1cm}\hfill
\includegraphics[height=3.9cm]{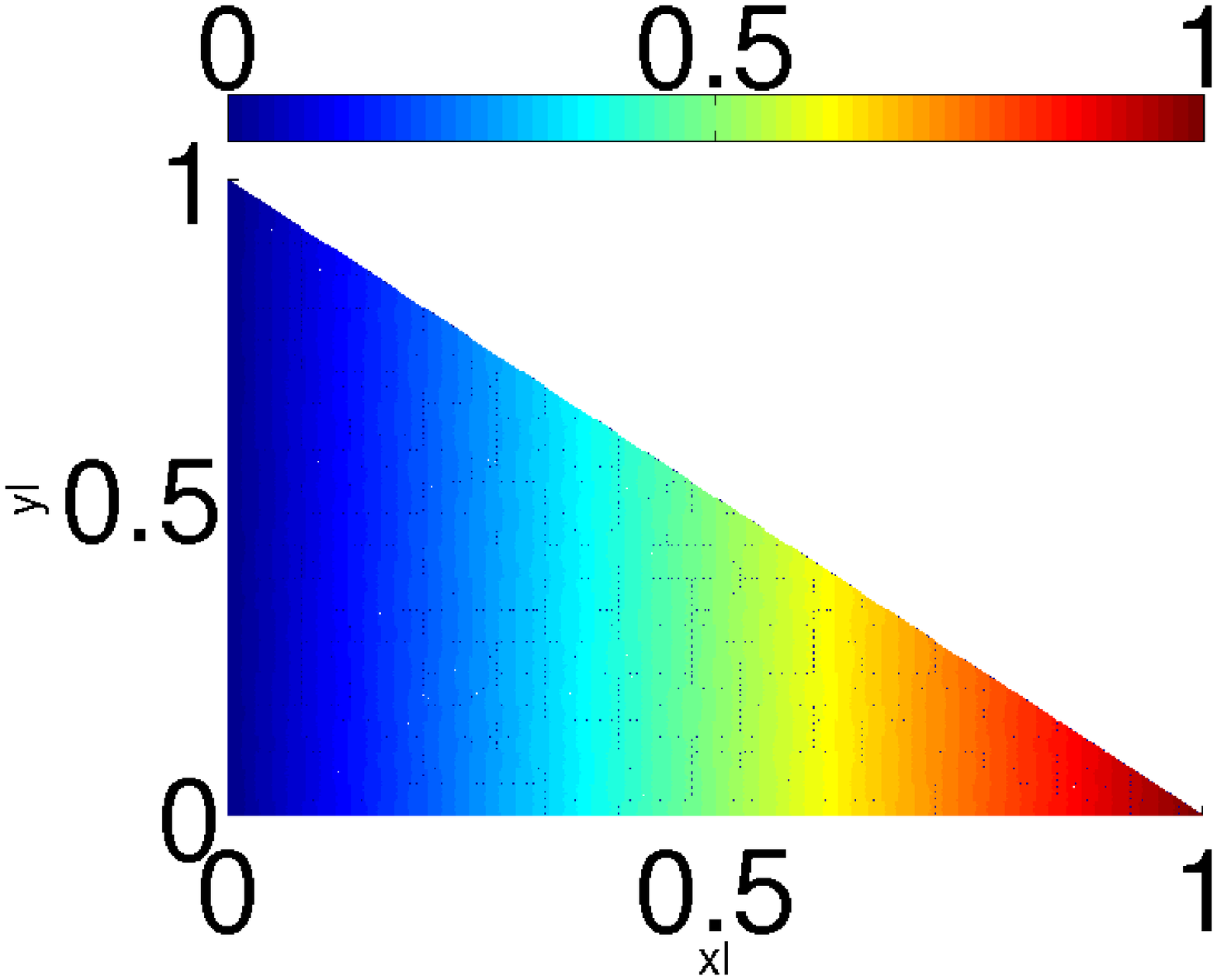}\hspace{0.1cm}\hfill
\includegraphics[height=3.9cm]{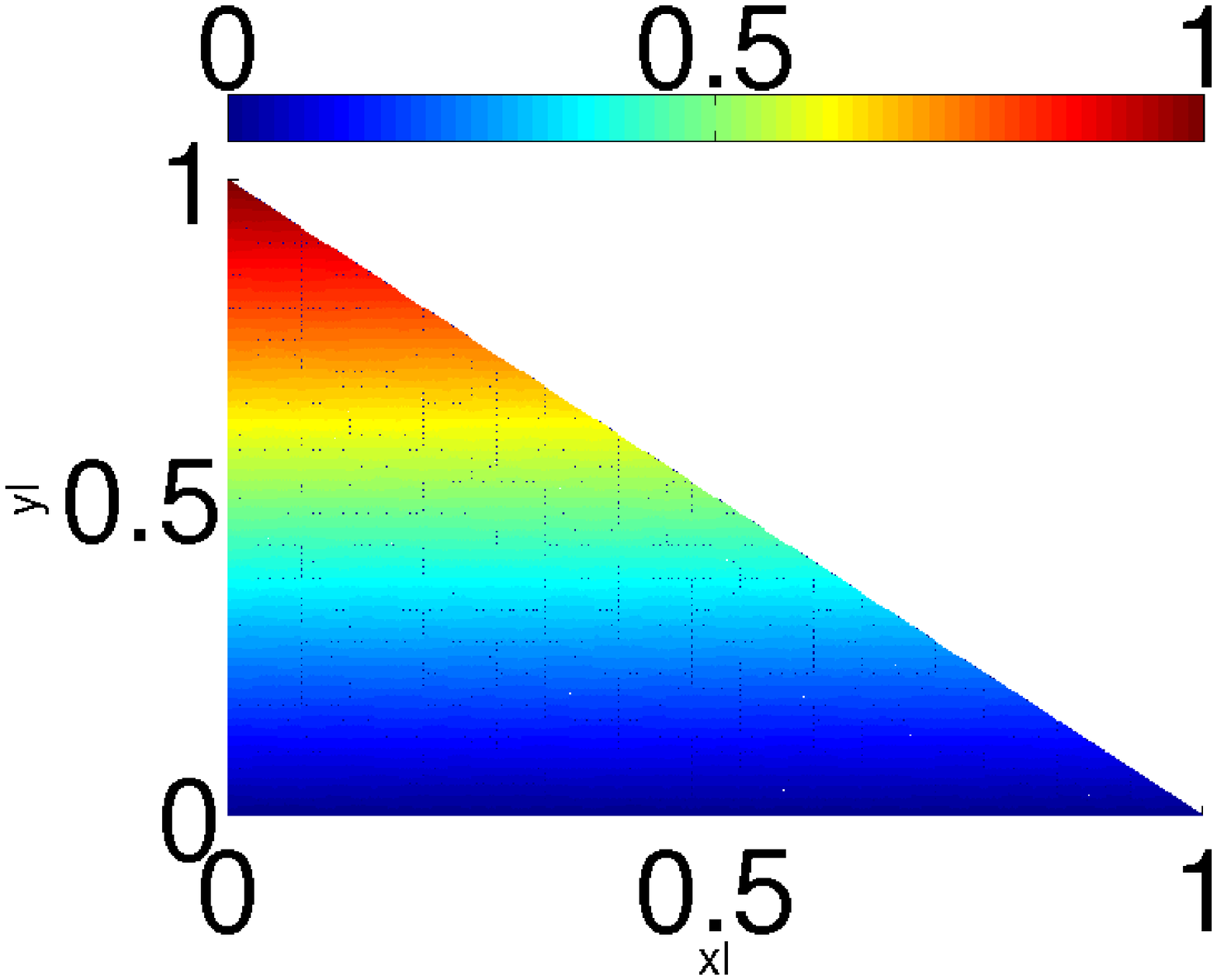}\hfill
\caption{\label{fig:scalar_1}Set of basis functions $\lambda_{i}$ of function space $P^{1}(K)$ (\ref{eq:h1Space}).}
\end{figure}
Fig. \ref{fig:scalar_1} shows the three basis functions $\lambda_{i}$ on a unit triangle. 

The basis $\{\lambda_{1},\lambda_{2},\lambda_{3}  \}$ is called nodal basis since the degrees of freedom are associated to the nodes of the patch. Furthermore each basis function $\lambda_{i}$ has the value $1$ at node $i$ of the patch and the value $0$ at all other nodes. The global ansatz functions of the finite element space $V_{h}$ are constructed from these local functions. Thereby several patches of the discretized domain can share the same node. On each of those patches we have a nodal basis function $\lambda_{i}$ which corresponds to the joint node. All of these basis functions then have the same degree of freedom. This means that a global ansatz functions consists of all local ansatz functions which have the value $1$ at the same node. The global ansatz functions are therefore globally continuous.

%This nodal basis is also used to construct the vectorial ansatz functions on the triangle. We rename it $\lambda_{i}=\varphi_{i}$. The elements of the nodal basis are shown in Fig. \ref{fig:scalar_1}. 

Next we want to construct vectorial finite elements of lowest order (but not constant) on the triangle which are $H(curl)$ conform. This means that the functions itself and the curl of the functions are quadratic Lebesgue integrable \cite{MON03}. As already mentioned it is important that the mathematical structure of the differential operators appearing in Maxwell's equations also holds for their discretized versions, i.e. the operators acting on our constructed function spaces of finite dimension. For the continuous operators we have the following important property: On simply connected subsets $\Omega$ of ${\mathbb R}^{2}$ the following exact sequence holds:
\begin{eqnarray}
  \label{eq:exSeq}
  H^{1}(\Omega)/{\mathbb R}\stackrel{\nabla}\longrightarrow H(curl,\Omega)\stackrel{\MyCurl}\longrightarrow L^{2}(\Omega)
\end{eqnarray}
where $L^{2}(\Omega)$ is the set of functions which are quadratic Lebesgue integrable on $\Omega$ and $H^{1}(\Omega)/ {\mathbb R}$ is the set of non constant functions in $H^{1}(\Omega)$. This sequence means: the operator $\nabla$ has an empty kernel on $H^{1}(\Omega)/ {\mathbb R}$. The range of $\nabla$ is a subset of $H(curl,\Omega)$ and it is exactly the kernel of $\MyCurl$. The range of $\MyCurl$ is the whole $L^{2}(\Omega)$. Hence we construct the local function spaces in such a way that the exact sequence:
\begin{eqnarray}
  \label{eq:exSeq1}
  W_{h}/ {\mathbb R}\stackrel{\nabla}\longrightarrow V_{h}\stackrel{\MyCurl}\longrightarrow S_{h}
\end{eqnarray}
holds, where 
\begin{eqnarray}
  \label{eq:subSpacesH}
W_{h}/ {\mathbb R}&\subset&H^{1}(\Omega)/ {\mathbb R},\\
V_{h}&\subset& H(curl,\Omega),\\
S_{h}&\subset& L^{2}(\Omega).
\end{eqnarray}
Constructing the linear scalar finite elements on a patch $K$ we already found:
\begin{eqnarray}
  \label{eq:spaces_Wh}
  W_{h}&=&\left\{ w\in H^{1}(\Omega):\;w|_{K}\in P^{1}(K) ,\forall K\right\}.
\end{eqnarray}
This can be seen when evaluating 
\begin{eqnarray}
  \label{eq:rangeCurlWh}
  \nabla (W_{h}/ {\mathbb R})|_{K}=\left\{\left(\begin{array}{c}a\\b\end{array}\right):\;a,b \in {\mathbb R} \right\}.
\end{eqnarray}
We have 
\begin{eqnarray}
  \label{eq:emptyKern}
  dim\left(W_{h}/ {\mathbb R}\right)=dim\left(\nabla (W_{h}/ {\mathbb R})\right)=2
\end{eqnarray}
hence $\nabla$ has an empty kernel on $W_{h}/ {\mathbb R}$. Furthermore the functions in $\nabla (W_{h}/ {\mathbb R})$ lie in the kernel of the curl operator. Now we want to construct $V_{h}$. The exact sequence for the discrete spaces (\ref{eq:exSeq1}) tells us that $\nabla (W_{h}/ {\mathbb R})\subset V_{h}$, which are the constant vectors. Since we wanted to have functions of lowest order but not only constant functions in $V_{h}$ we have to extend it. Again the exact sequence tells us how to make this extension:
\begin{eqnarray}
  \label{eq:exactVhconst}
  \MyCurl V_{h}&=&S_{h}\subset L^{2}(\Omega)
\end{eqnarray}
With only constant elements in $V_{h}$ it follows that $S_{h}=\{ 0\}$. Let us extend $V_{h}$ in such a way that $S_{h}$ includes at least the constant functions $S_{h}=\left\{ s\in L^{2}(\Omega)/ {\mathbb R}:\;s|_{K}\in P^{0}(K) \right\}$. The vectorial functions in $V_{h}$ which we include in addition to the constant functions are polynomials of $x$ and $y$. In the $x$-component the $y$-variable is not allowed to have a degree higher than one and vice versa. Otherwise the curl of the vector would not be constant. Therefore we use the following ansatz:
\begin{eqnarray}
  \label{eq:findV}
  v=\left( \begin{array}{c}p_{1}(x)+by\\cx+p_{2}(y)\end{array} \right)\in V_{h}\Rightarrow \MyCurl v=c-b,
\end{eqnarray}
where $p_{1}$, $p_{2}$ are arbitrary non constant polynomials. Since the exact sequence tells us that only elements from $\nabla (W_{h}/ {\mathbb R})$ lie in the kernel of $\MyCurl$ we can deduce $p_{1}=p_{2}=0$: vectors of the form $\left( \begin{array}{c}p_{1}(x)\\p_{2}(y)\end{array} \right)$ also lie in the kernel of $\MyCurl$ but are not elements of $\nabla (W_{h}/ {\mathbb R})$. Since we wanted to extend $V_{h}$ in a minimal way we furthermore choose $c=-b$. We found:
\begin{eqnarray}
  \label{eq:NedI}
  V_{h}=N_{0}^{I}(K)=\left\{v=\left(\begin{array}{c}a_{x}\\a_{y}\end{array}\right) +b  \left(\begin{array}{c}y\\-x\end{array}\right)\,,\;a_{x},a_{y},b \in {\mathbb R} \right\},\;dim(N_{0}^{I}(K))=3.
\end{eqnarray}
These elements of lowest order were discovered independently by a number of authors, see \cite{MON03}. The above considerations can be extended to higher order elements as well and were first constructed by Nedelec \cite{NED80}.

\begin{figure}[h]
\psfrag{xl}{$x$}
\psfrag{yl}{$y$}
\psfrag{0}{$0$}
\psfrag{1}{$1$}
\psfrag{0.5}{$0.5$}
(a)\hspace{4.9cm}(b)\hspace{4.7cm}(c)\\
\includegraphics[height=4.7cm]{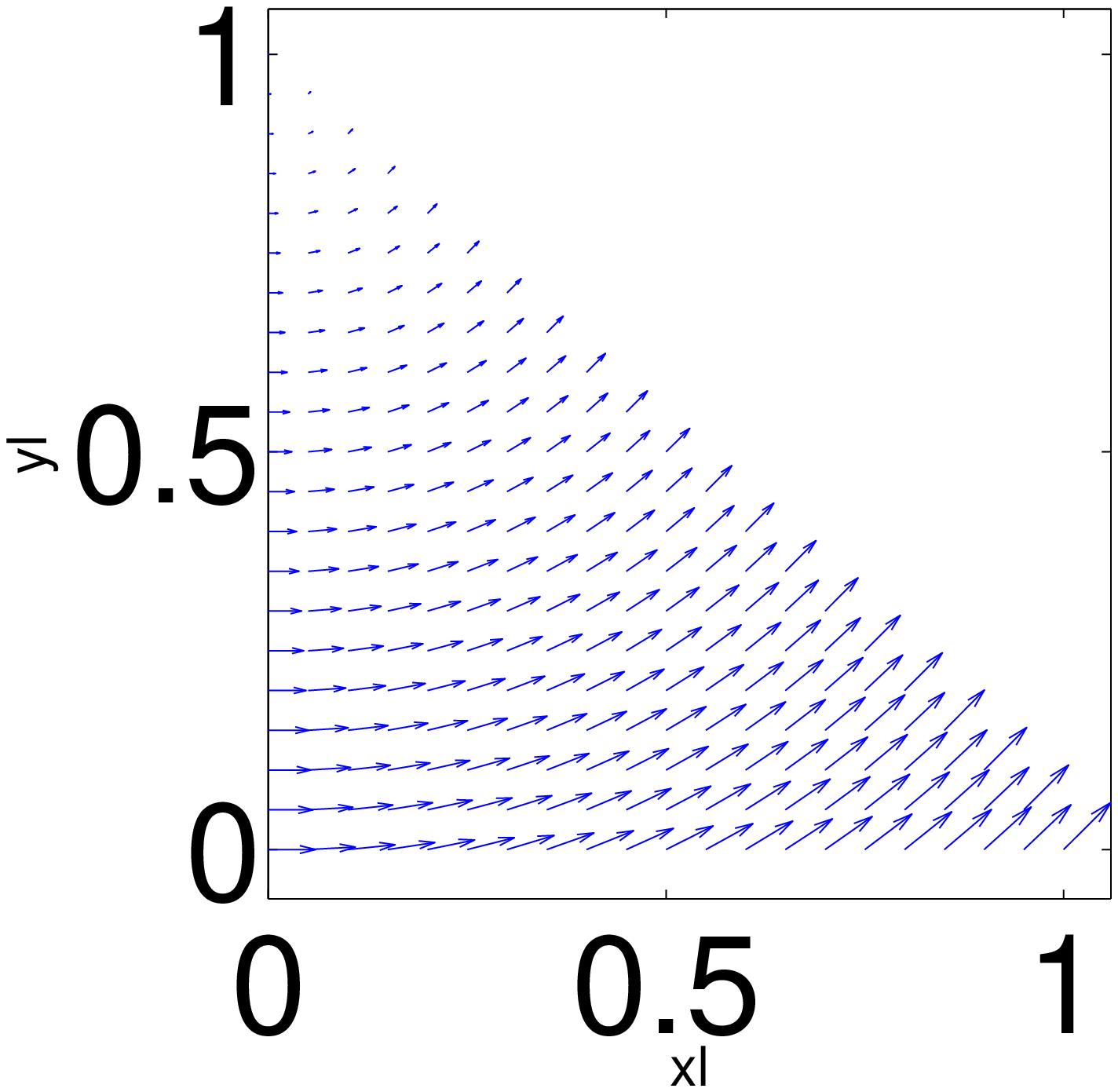}\hspace{0.0cm}\hfill
\includegraphics[height=4.7cm]{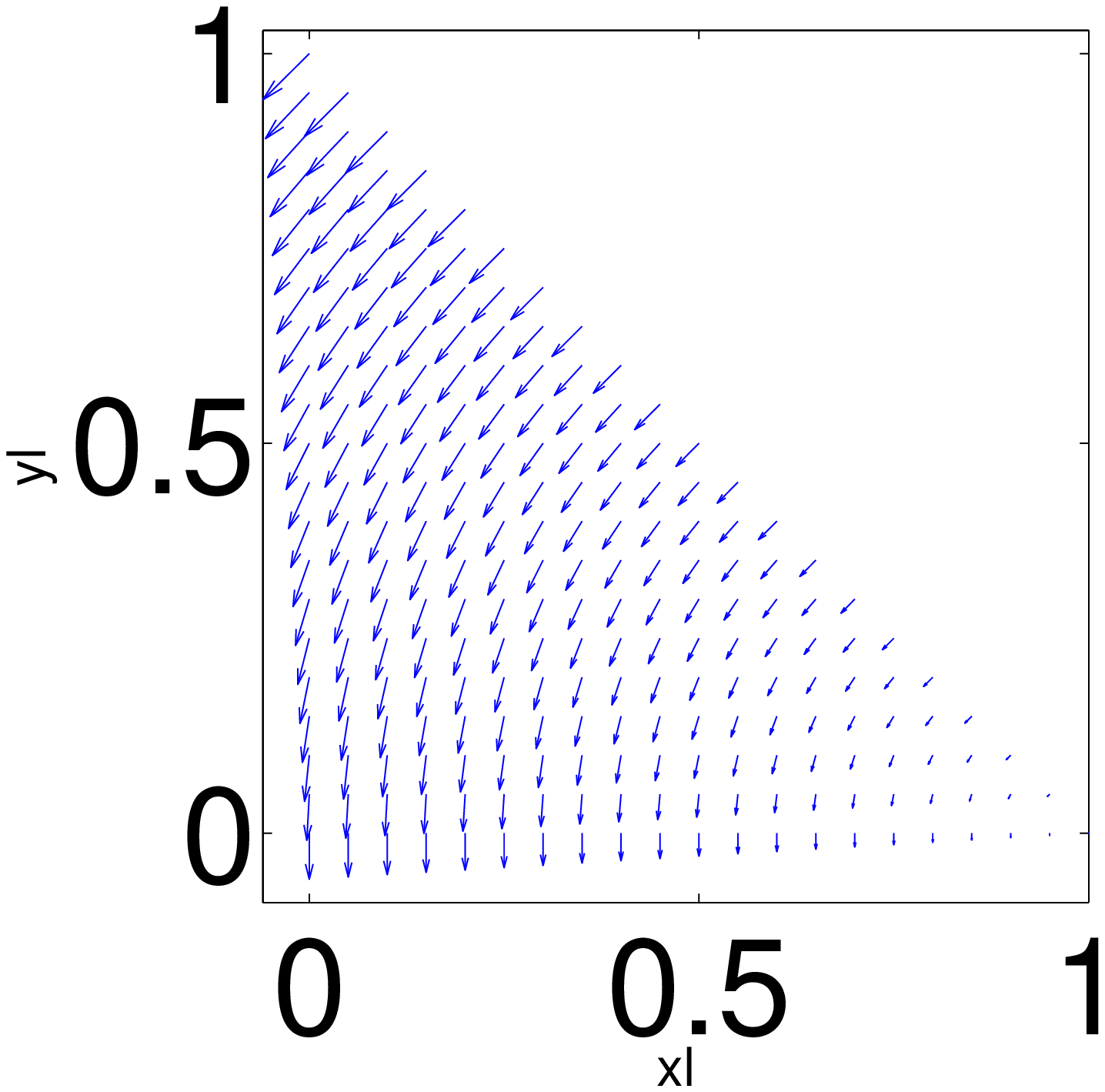}\hspace{0.0cm}\hfill
\includegraphics[height=4.7cm]{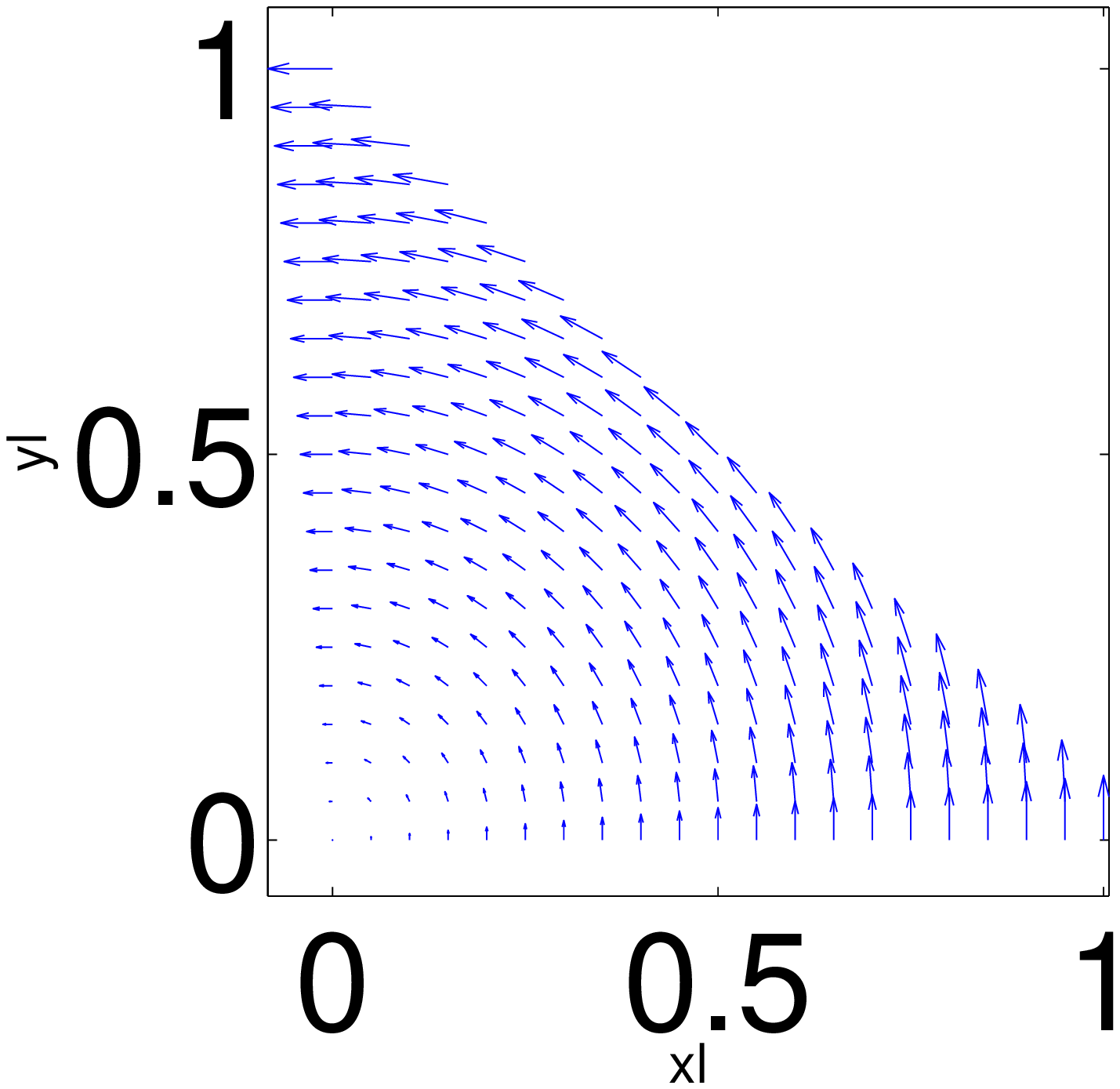}\hfill
\caption{\label{fig:Hcurl}Set of basis functions $\varphi_{ij}$ of function space $N_{0}^{I}(K)$, Eq. (\ref{eq:NedI}).}
\end{figure}

After finding the local function space we define the degrees of freedom. We associate them to the three edges connecting the nodes $1\rightarrow 2,2\rightarrow 3,3\rightarrow 1$ of the patch:
\begin{eqnarray}
  \label{eq:Ndof}
  \psi_{12}(\MyField{v})&=&\int\limits_{12}\MyField{v}\cdot\MyField{\tau}ds,\\
  \psi_{23}(\MyField{v})&=&\int\limits_{23}\MyField{v}\cdot\MyField{\tau}ds,\\
  \psi_{31}(\MyField{v})&=&\int\limits_{31}\MyField{v}\cdot\MyField{\tau}ds,
\end{eqnarray}
where the integral $\int\limits_{ij}$ is carried out along the edge $ij$ which starts at node $i$ and ends at node $j$. The quantity $\tau$ is the tangential vector of the edge. The degrees of freedom therefore correspond to the integral of the tangential component along the edges of the patches. The basis functions which we construct using (\ref{eq:deg}) are therefore associated to the edges of the patch and the finite elements are called edge elements:
\begin{eqnarray}
  \label{eq:Nbasis}
  \psi_{12}(\varphi_{12})&=&1\,=\int\limits_{12}\left(   \left(\begin{array}{c}a_{x,1}\\a_{y,1}\end{array}\right) +b_{1}  \left(\begin{array}{c}y\\-x\end{array}\right)  \right)\cdot\MyField{\tau}ds,\nonumber\\
  \psi_{23}(\varphi_{12})&=&0\,=\int\limits_{23}\left(   \left(\begin{array}{c}a_{x,2}\\a_{y,2}\end{array}\right) +b_{2}  \left(\begin{array}{c}y\\-x\end{array}\right)  \right)\cdot\MyField{\tau}ds,\nonumber\\
  \psi_{31}(\varphi_{12})&=&0\,=\int\limits_{31}\left(   \left(\begin{array}{c}a_{x,3}\\a_{y,3}\end{array}\right) +b_{3}  \left(\begin{array}{c}y\\-x\end{array}\right)  \right)\cdot\MyField{\tau}ds.\nonumber
\end{eqnarray}
Carrying out the integrals for all basis functions we can determine all unknown coefficients. The resulting basis functions are \cite{MON03}:
\begin{eqnarray}
  \label{eq:NBasis1}
  \MyField{\varphi}_{12}&=&\lambda_{1}\nabla\lambda_{2}-\lambda_{2}\nabla\lambda_{1},\nonumber\\
  \MyField{\varphi}_{23}&=&\lambda_{2}\nabla\lambda_{3}-\lambda_{3}\nabla\lambda_{2},\nonumber\\
  \MyField{\varphi}_{31}&=&\lambda_{3}\nabla\lambda_{1}-\lambda_{1}\nabla\lambda_{3},\nonumber
\end{eqnarray}
where $\lambda_{i}$ are the nodal basis functions. The basis $\{\varphi_{12},\varphi_{23},\varphi_{31}  \}$ is shown in Fig. \ref{fig:Hcurl}.
When we construct global basis functions from the local functions $\varphi_{ij}$ then the local functions of neighboring patches which correspond to the joint edge share the same degree of freedom. Since the degree of freedom is defined via the integral of the tangential component this means that the global ansatz functions have a continuous tangential component. The discretization of the computational domain is performed respecting the material boundaries (i.e. all material boundaries lie on edges of the discretization). Therefore the finite element solution generically includes the physical property of electric fields having a continuous tangential component across boundaries. The normal component can be discontinuous in general.

\section{Transparent boundary conditions}
\label{sec:Transparent}
\begin{figure}[h]
\centering
\includegraphics[height=6cm]{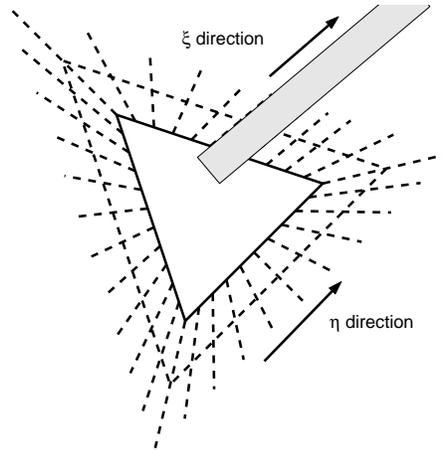}\hfill
\caption{Prismatoidal coordinate system $(\xi,\eta)$ introduced in the exterior domain. The waveguide structure yields solutions analytic in $\xi$-direction.}
\label{Fig:prismatoidal_coord}
\end{figure}

In this section we present our realization of transparent boundary conditions \cite{Zschiedrich03}. We use the perfectly matched layer (PML) method \cite{BerPML} and construct the Dirichlet to Neumann operator mentioned in Sec. \ref{sec:Propagation}. For simplicity and the sake of a clear presentation we restrict ourselves to the 2D case. The ideas carry over to the 3D case of Maxwell's equations. The computational domain has to be polygonal and star-shaped. A certain class of inhomogeneities in the exterior is allowed which also covers open waveguide structures playing an essential role in integrated optics, see Fig. \ref{Fig:prismatoidal_coord}. Here a computational domain with a waveguide structure is shown. The Silver-M\"uller radiation condition (\ref{eq:smmw1}) does not hold true for such inhomogeneous domains and has to be generalized. The pole condition was introduced in \cite{Schmidt02H} as a general concept to define radiation conditions for wave propagation problems. This also gave new insight into the PML method.
The implementation is based on prismatoidal coordinate systems, shown in Fig. \ref{Fig:prismatoidal_coord}. New coordinates $\xi$ and $\eta$ are introduced, where $\xi$ denotes a generalized distance variable. The central idea is to decompose the exterior domain $\Omega_{ext}^{(x,y)}$ into a finite number of segments \ref{fig:pml_geometry2D}. Each of the segments carries its own coordinate system such that a global distance variable $\xi$ can be introduced, Fig. \ref{fig:pml_geometry2D}. Maxwell's equations are then semi-discretized in a generalized angular variable $\eta$ and the PML method is realized via the complex continuation along the $\xi$-direction. The decomposition of the exterior is based on straight non-intersecting rays which connect each vertex of the polygonal boundary $\partial\Omega$ with infinity, shown in Fig. \ref{fig:pml_geometry2D}. The arising semi-infinite quadrilaterals $Q_{j}^{(x,y)}$ with their $xy$-coordinate system can then be given as the range of reference rectangles $Q_{j}^{(\xi,\eta)}$ under a local bilinear transformation $B_{j}^{loc}$, Fig \ref{fig:pml_geometry2D}:
\begin{figure}[t]
\includegraphics[width=12cm]{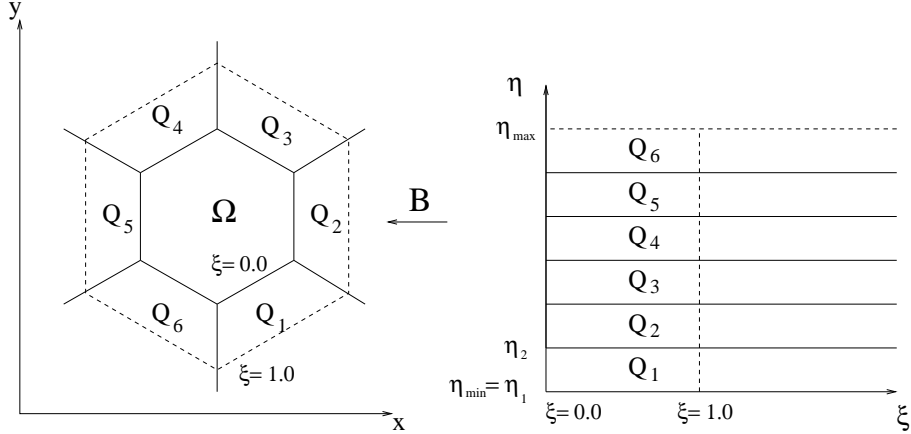}
\caption{Prismatoidal coordinate system. Each segment $Q_j$ is the image of
a reference element under a bilinear mapping $B^{loc}_j$. These local mappings
are combined to a global mapping $B$ which is continuous in $\eta$.}
\label{fig:pml_geometry2D}
\end{figure}
\begin{eqnarray}
  \label{eq:Bloc}
  B_{j}^{loc}&:&Q_{j}^{(\xi,\eta)}\rightarrow Q_{j}^{(x,y)}.
\end{eqnarray}
We then have a transformation of the semi-infinite rectangle $Q_{j}^{(\xi,\eta)}:=[0,\infty]\times[\eta_{j},\eta_{j+1}]$ onto $Q_{j}^{(x,y)}$ such that lines with $\xi = const$ remain parallel under $B_{j}^{loc}$. The exact mathematical definitions for the coordinate system, radial and normal rays are given in \cite{Zschiedrich03}.

Now Maxwell's equations are formulated in the $\xi\eta$ coordinate system:
\begin{eqnarray}
  \label{eq:MXxieta}
  \MWStar{(\xi,\eta)},
\end{eqnarray}
with transformed permittivity and permeability:
\begin{eqnarray}
  \label{eq:epsStar}
\eps_{*}&=&|J|J^{-1}\eps J^{-T} \\
{\mu_{*}}^{-1}&=&\frac{1}{|J|}J^{T}\muInv J,
\end{eqnarray}
where $J$ denotes the Jacobian of the coordinate transformation $B_{j}^{loc}$ and $|J|$ its determinant. In order to formulate the semi-discrete formulation of Maxwell's equation one first has to derive the weak formulation of Maxwell's equations. Therefore (\ref{eq:MXxieta}) is multiplied with a test function $\MyField{w}$ and integrated along $\Gamma=\partial\Omega$. The rigorous mathematical definitions of the involved function spaces is beyond the scope of this paper and can be found in \cite{Zschiedrich03}.
%\begin{eqnarray}
%  \label{eq:formsout}
%a(\MyField w,\MyField v)&=&\int_{\Gamma}\MyField w\cdot\left(\nabla_{\xi\eta}\times\frac{1}{\mu_{*}}\nabla_{\xi\eta}\times\MyField{v}-\cdot\omega^{2}\eps_{*}\MyField{v}\right)d\eta\\
%\end{eqnarray}
%The problem to solve in the exterior domain than reads:\\
%Find $\MyField{v}\in W^{2}(\Omega_{ext}^{(\xi,\eta)})$ such that for all $\MyField{w}\in H_{\pi}(curl,[\eta_{min},\eta_{max}])$ and all $\xi \in {\mathrm{R}}_{+}$
%\begin{eqnarray}
%  \label{eq:weakPML}
%\end{eqnarray}
%where $H_{\pi}(curl,[\eta_{min},\eta_{max}])$ is the space of periodic functions in $H(curl,[\eta_{min},\eta_{max}])$ and
%\begin{eqnarray}
%  \label{eq:defW2}
%  W^{2}(\Omega_{ext}^{(\xi,\eta)})=\left\{
%    \begin{array}[h]{c}
%v(\xi_{0},\eta)\in H_{\pi}(curl,[\eta_{min},\eta_{max}]):\forall \xi_{0} \in {\mathrm{R}}_{+}\\
%v(\xi,\eta_{0})\in C^{2}({\mathrm{R}}_{+}):\forall \eta \in [\eta_{min},\eta_{max}]
%    \end{array}
%    \right.
%\end{eqnarray}
The semi discretization with respect to $\eta$ of the scattered outgoing electric field is performed via the ansatz:
\begin{eqnarray}
  \label{eq:pmlAnsatz}
  \MyField{E}^{h}_{out}(\xi,\eta)=\sum\limits_{j=1}^{N_{B}}E^{h}_{out,j}(\xi)\psi_{j}(\eta)
\end{eqnarray}
where the functions $\{\psi_{1},\dots,\psi_{N_{B}}\}$ form a basis of $S_{h}$ which is the trace space of the finite element space $V_{h}$ of the interior domain. This means that each $\psi_{i}$ can be found in the set of basis functions in $V_{h}$ restricted to the boundary $\Gamma$. Using ansatz (\ref{eq:pmlAnsatz}) in the weak formulation of Maxwell's equations gives us a linear system of differential equations for the coefficient vector $\MyField E^{h}_{out}(\xi)=(E^{h}_{out,1}(\xi),\dots,E^{h}_{out,N_{B}}(\xi))$. We will denote it by:
\begin{eqnarray}
  \label{eq:pmlSemi}
  A(\xi)^{h}_{out}\MyField E^{h}_{out}(\xi)=0,
\end{eqnarray}
where $A(\xi)^{h}_{out}$ is a differential operator acting on the coefficient vector $\MyField E^{h}_{out}(\xi)$.
The PML method is realized by replacing $\xi\rightarrow \gamma\xi$ ($\Re({\gamma})>0$, $\Im({\gamma})>0$), which is the complex continuation of the exterior solution. Furthermore the unbounded domain $\Omega_{ext}^{(\xi,\eta)}$ is replaced with the bounded domain $\Omega_{PML}:=\{ (\xi,\eta)\in \Omega_{ext}^{(\xi,\eta)}:\xi\in[0,\rho];\eta\in[\eta_{min},\eta_{max}] \}$. Because of the expected absorbing character of the PML we impose zero Dirichlet boundary conditions on the outer boundary $\xi=\rho$. The resulting PML system then reads
\begin{eqnarray}
  \label{eq:PMLsystem}
  A(\gamma\xi)^{h}_{out}\MyField E^{h}_{PML}(\xi)=0,
\end{eqnarray}
with $E^{h}_{PML}(\xi)=E^{h}_{out}(\gamma\xi)$. Finally the PML system (\ref{eq:PMLsystem}) is discretized in the $\xi$ variable with the finite element method.

The complete discretization of the exterior domain may then be interpreted as a FEM discretization on quadrilaterals. Their quality depends on the initial choice of the rays. The solution in the exterior is analytic in $\xi$ direction. It is therefore advantageous to choose high order finite elements for the $\xi$-discretization.
%, i.e. each component of the exterior solution $E^{h}_{PML}(\xi)=(E^{h}_{PML,1}(\xi),\dots,E^{h}_{PML,N_{B}}(\xi))$ is approximated by:
%\begin{eqnarray}
%  \label{eq:PMLdisc}
% E^{h}_{PML,i}(\xi)=\sum\limits_{n=1}^{N_{\xi}}c_{PML,i,n}\Phi_{n}(\xi),
%\end{eqnarray}
%where $\{\Phi_{1},\dots,\Phi_{N_{\xi}}$ is a basis of the finite element space

\section{Application: Optimization of photonic crystal fiber design}
\label{sec:Application}
\begin{figure}[ht]
(a)\hspace{4.8cm}(b)\hspace{4.8cm}(c)\\
\includegraphics[width=4.5cm,height=4.5cm]{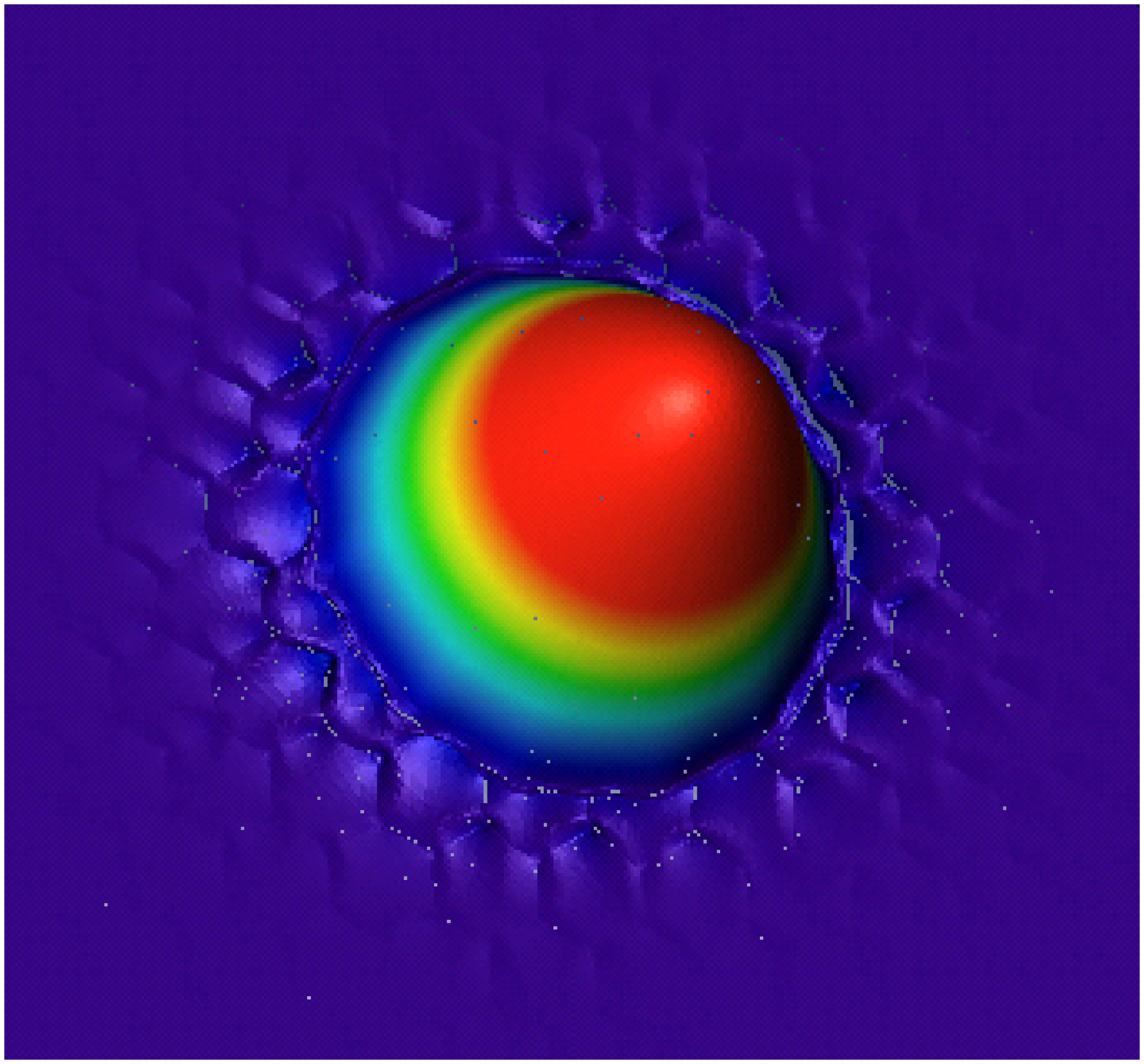}\hfill
\includegraphics[width=4.5cm,height=4.5cm]{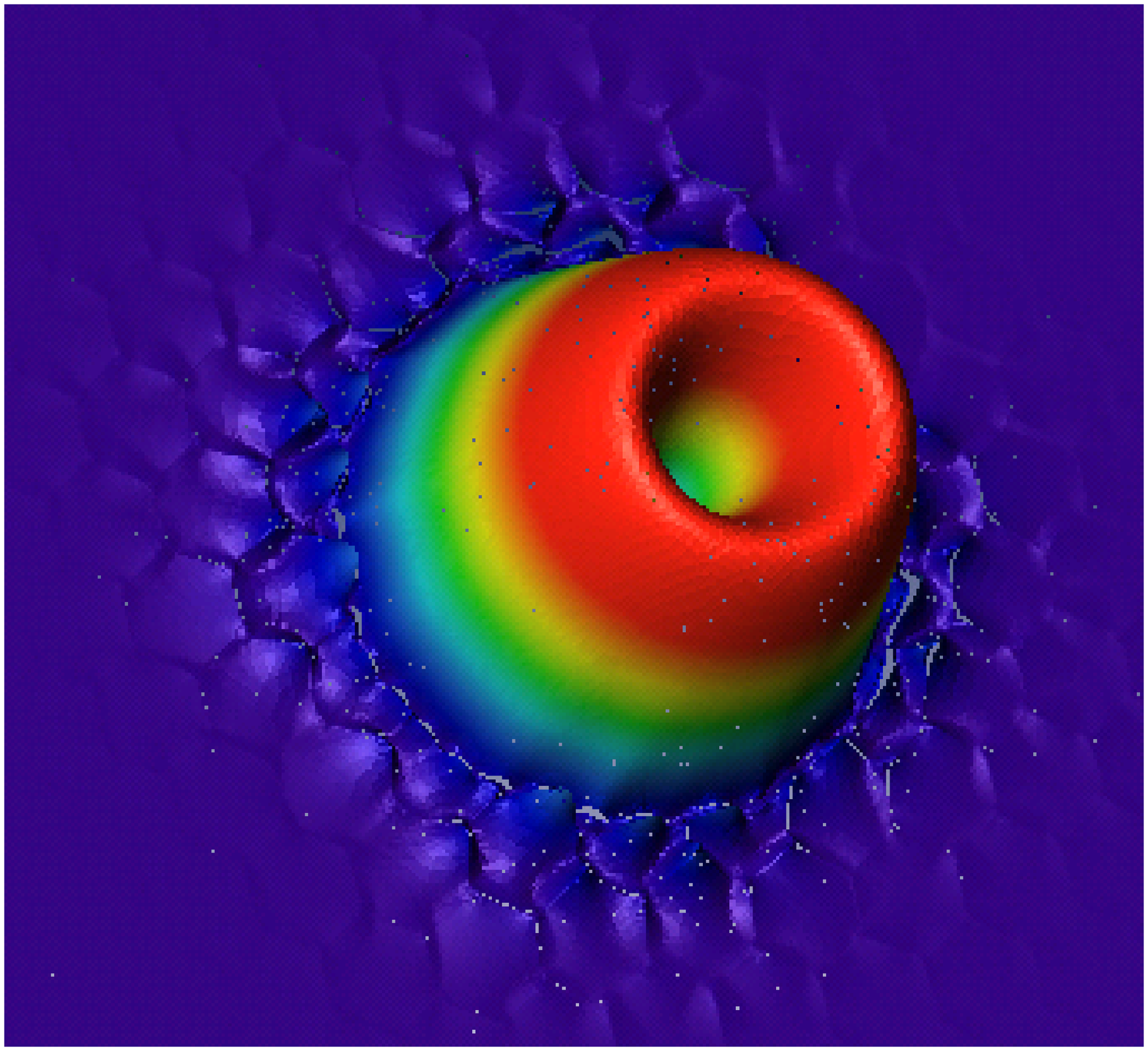}\hfill
\includegraphics[width=4.5cm,height=4.5cm]{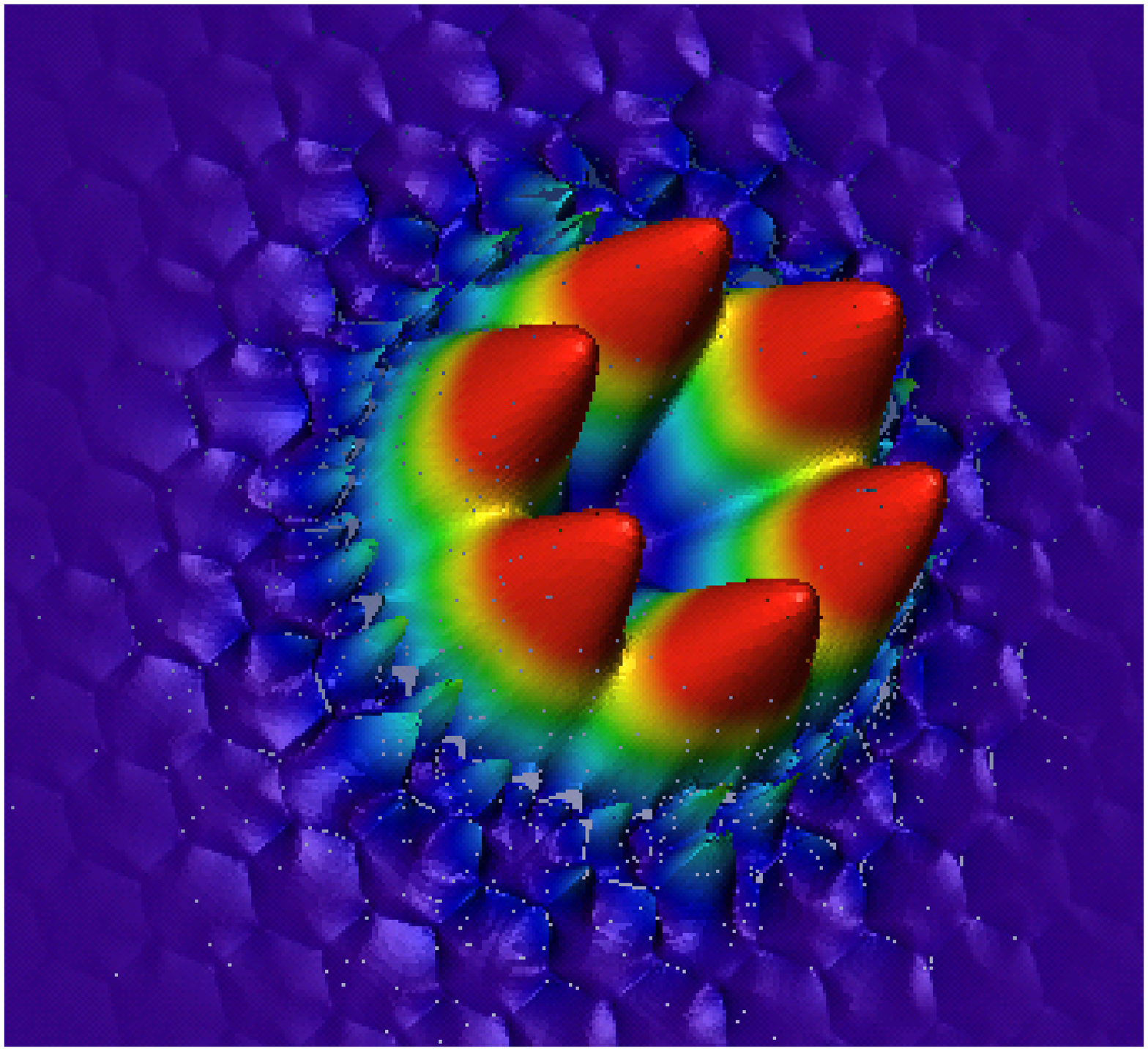}\hfill
\caption{\label{fig:coreModes}First, second and fourth fundamental core modes of HCPCF illustrated in Fig. \ref{fig:fem}(a) - Parameters: $\Lambda=1550\,$nm, $r=300\,$nm, $w=50\,$nm, $t=170\,$nm, $6$ cladding rings, $\lambda=589\,$nm, see Fig. \ref{fig:hcpcfTriang}(a) for definition of parameters.}
\end{figure}
In the last section we want to apply the finite element method to the computation of propagating modes in hollow core photonic crystal fibers (HCPCFs) \cite{CRE99,RUS03,COU06}. The results which are presented here are a summary of our work published in \cite{POM07,POM07a,ZSC07}. The computational domain and triangulation of a HCPCF was already shown in Fig. \ref{fig:fem}(a) and (b). The mathematical formulation for this problem type was given in Sec. \ref{sec:Propagation}. Here we are interested in radiation losses from photonic crystal fibers which means that we compute leaky propagating modes . Therefore we have to take the exterior of the fiber into account and apply transparent boundary conditions to the computational domain. Fig. \ref{fig:coreModes} shows the first, second, and fourth fundamental leaky core modes. We approximated the exterior of the computational domain by a glass cladding of infinite size. This approximation is justified if no light which is leaving the microstructured core of a HCPCF is reflected back from the outside of the cladding.
\begin{figure}[ht]
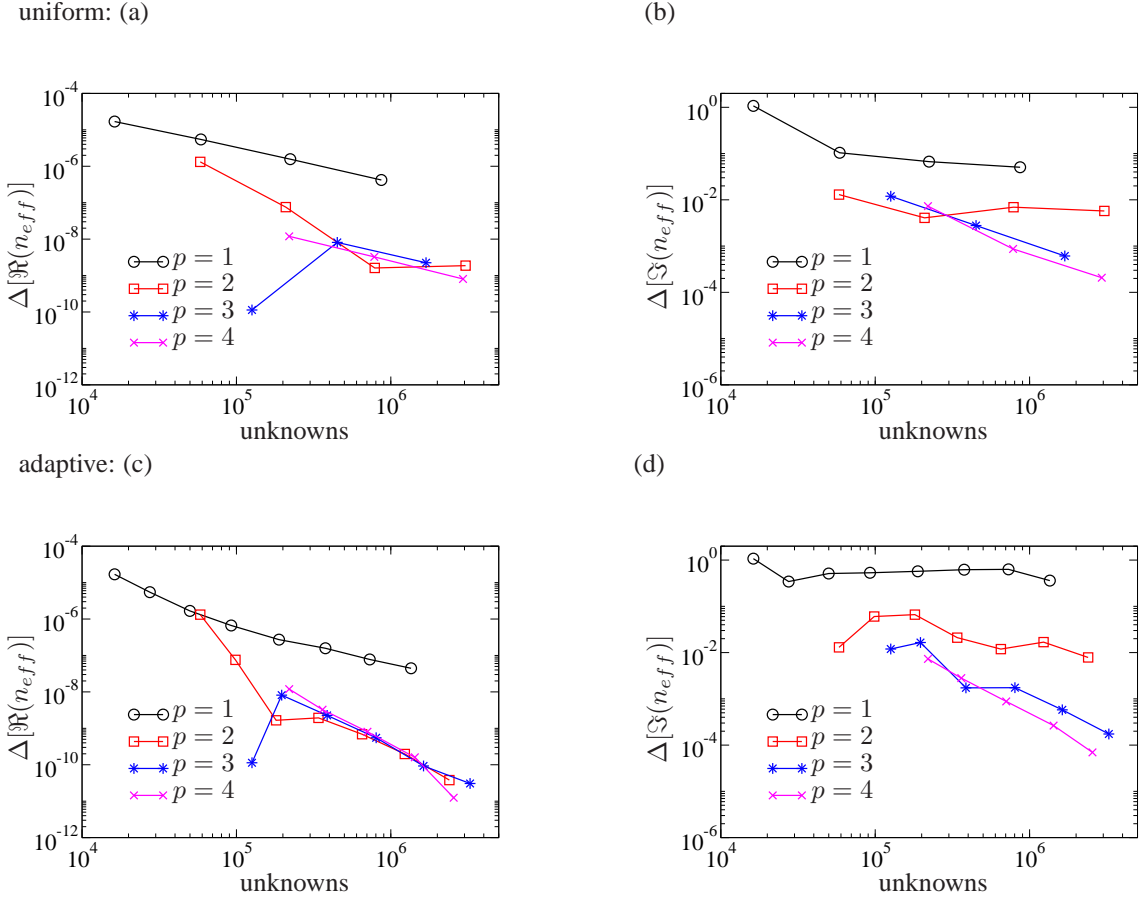

\psfrag{deltare}{$\Delta[\Re(n_{eff})]$}
\psfrag{deltaim}{$\Delta[\Im(n_{eff})]$}
\psfrag{unk}{unknowns}
\psfrag{p=1}{$p=1$}
\psfrag{p=2}{$p=2$}
\psfrag{p=3}{$p=3$}
\psfrag{p=4}{$p=4$}
\psfrag{no}{uniform}
\psfrag{go}{strategy B}
\psfrag{ad}{strategy A}
uniform: (a) \hspace{6.5cm}(b)\vspace{7mm}\\
\includegraphics[width=6.4cm]{fig/convSFBrealNo.eps}\hfill
\includegraphics[width=6.4cm]{fig/convSFBimagNo.eps}\hfill\\
adaptive: (c)\hspace{6.4cm}(d)\vspace{7mm}\\
\includegraphics[width=6.4cm]{fig/convSFBrealAd.eps}\hfill
\includegraphics[width=6.4cm]{fig/convSFBimagAd.eps}\hfill\\
\caption{\label{fig:hcpcfConv}Relative error of fundamental eigenvalue in dependence on number of unknowns of FEM computation for uniform and adaptive refinement strategies and finite element degrees $p$. Parameters: $\Lambda=1550\,$nm, $r=300\,$nm, $w=50\,$nm, $t=170\,$nm, 6 cladding rings, wavelength $\lambda=589\,$nm.}
\end{figure}
\begin{figure}[ht]
\psfrag{neffimag}{$\Im(n_{\mathrm{eff}})$}
\psfrag{lambda}{$\Lambda$}
\psfrag{wlabel}{$w$}
\psfrag{core}{$t$}
\psfrag{nrows}{cladding rings}
\psfrag{hole}{$r$}
(a)\hspace{8.5cm}(b)\hspace{5.5cm}\\
\includegraphics[width=6cm]{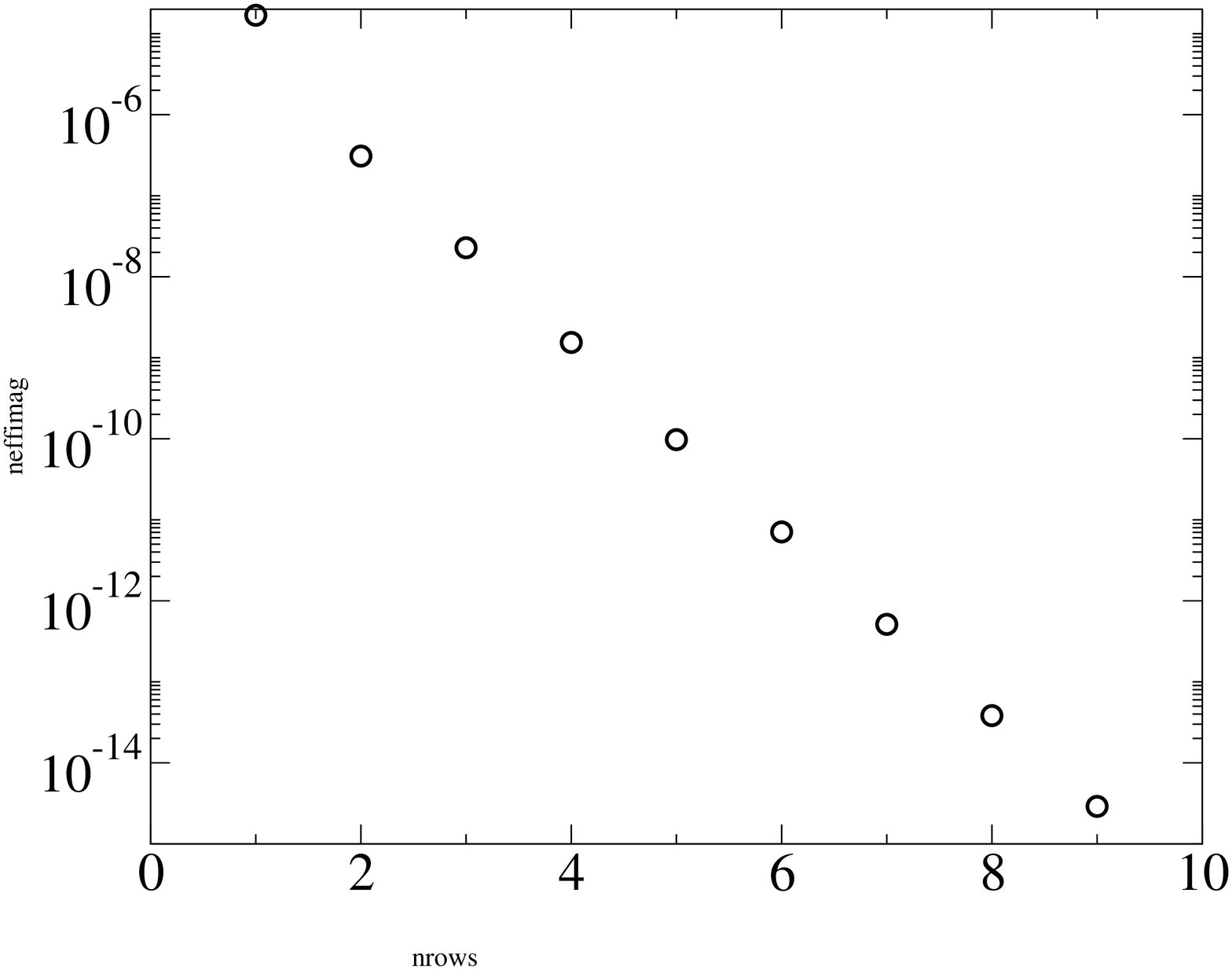}\hfill
\includegraphics[width=6cm]{fig/pitchScan2_.eps}\hfill\\
(c)\hspace{8.5cm}(d)\hspace{5.5cm}\\
\includegraphics[width=6cm]{fig/coreSurroundScan_.eps}\hfill
\includegraphics[width=6cm]{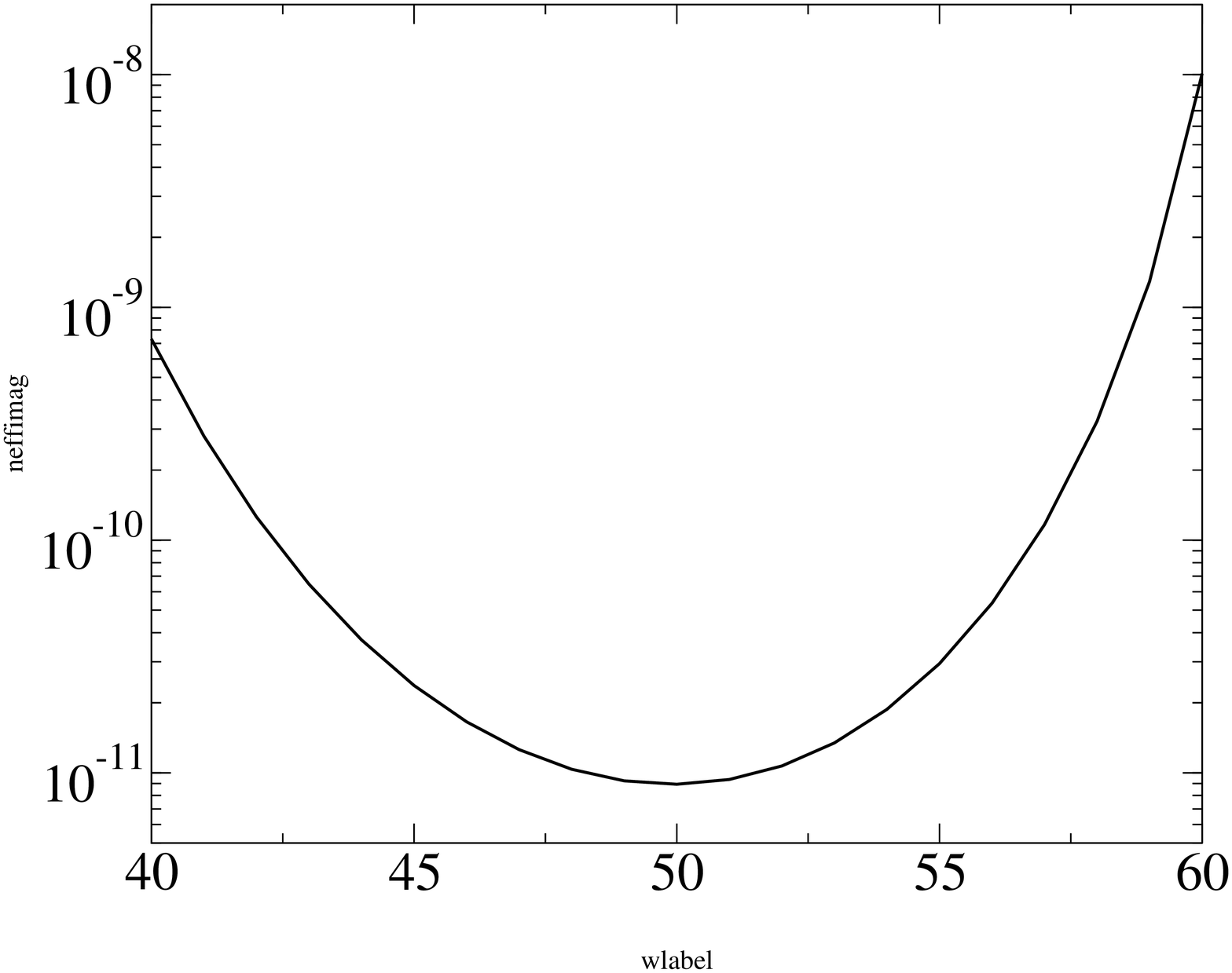}\hfill\\
(e)\\
\includegraphics[width=6cm]{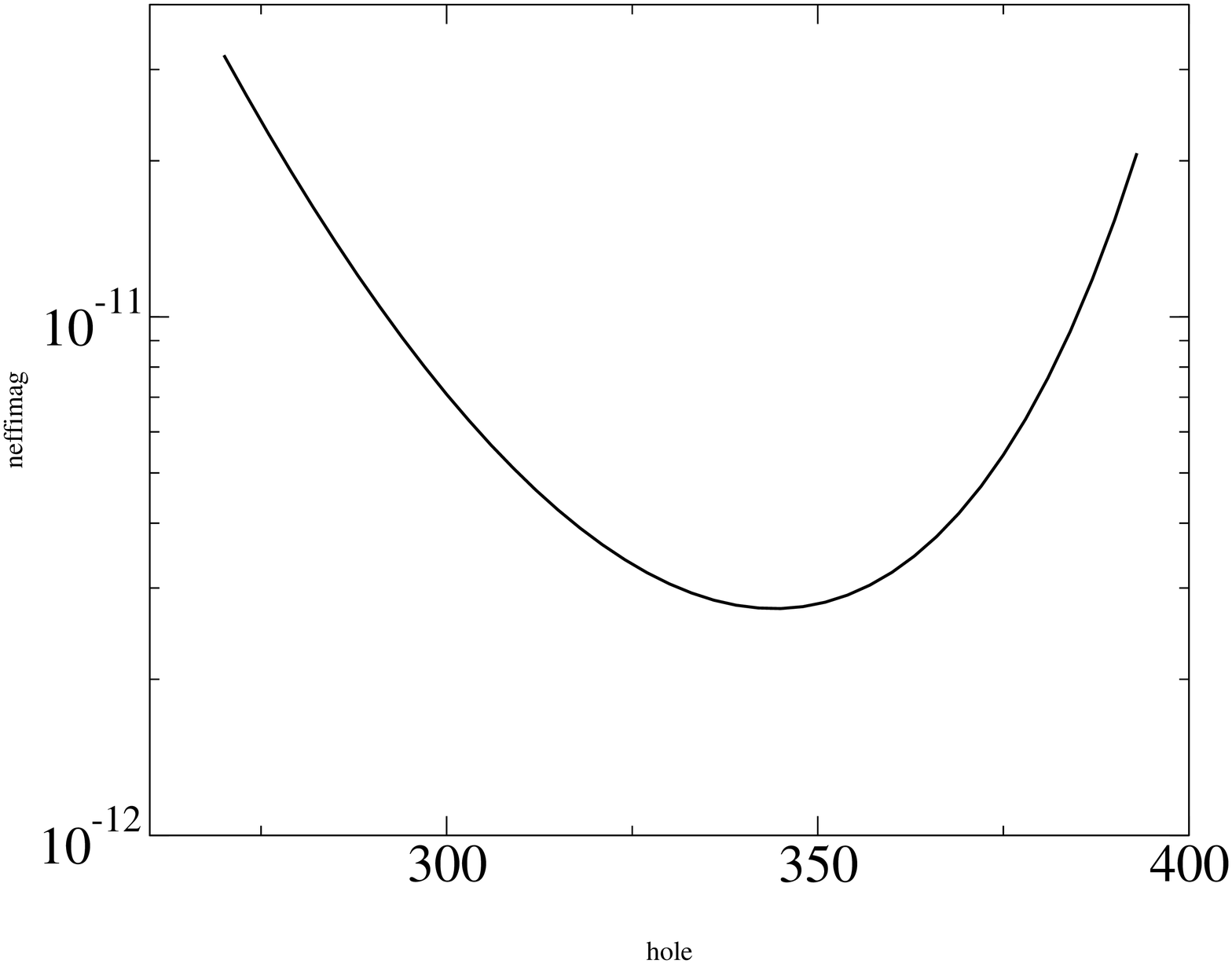}
\caption{\label{fig:geoScan}Imaginary part of effective refractive index $\Im(n_{\mathrm{eff}})$ in dependence on: (a) number of cladding rings, (b) pitch $\Lambda$, (c) core surround thickness $t$, (d) strut thickness $w$, (e) hole edge radius $r$. Parameters: $\Lambda=1550\,$nm, $r=300\,$nm, $w=50\,$nm, $t=170\,$nm, 6 cladding rings, wavelength $\lambda=589\,$nm.}
\end{figure}
When applying transparent boundary conditions to the computational domain the propagation constant $k_z$ (\ref{eq:evp}) becomes complex \cite{Zschiedrich2005a}. The corresponding leaky mode is therefore exponentially damped according to ${\mathrm{exp}}({-\Im(k_{z})z})$ while propagating along the fiber. It can be shown from Maxwell's equations that the imaginary part of $k_z$ is proportional to the power flux of the electric field across the boundary $\Gamma$ of the computational domain \cite{ZSC07}. This is a quantity one often wants to minimize in application. We will quantify the radiation losses by the imaginary part of the effective refractive index (\ref{eq:neff}). First we are interested how accurately we can compute the eigenvalues. Fig. \ref{fig:hcpcfConv} shows the convergence of the real and imaginary part of the effective refractive index for an uniform and adaptive grid refinement strategy and different finite element degrees \cite{POM07a}. For both refinement strategies the real part converges very fast the imaginary part is much harder to compute \cite{Becker:01a}. We see that high order finite elements are needed to get an accurate solution for the imaginary part. In an uniform refinement step each triangle is subdivided into four smaller ones. This allows to compute the FEM solution more accurately and the relative error of the real and imaginary part decreases, see Fig. \ref{fig:hcpcfConv}(a), (b). On the other hand the number of unknowns and therewith the computational time and memory requirements increase. However if one utilizes the sparsity of the linear equation system (\ref{eq:MWweak4}) which has to be solved the memory and computational time scale linearly with the number of unknowns \cite{HOL06}. The eigenvalue obtained from the most accurate FEM computation is used as the reference solution for the convergence plots. Since for the finite element method convergence is proven mathematically it is reasonable to assume that the FEM solution converges towards the exact continuous solution. Fig. \ref{fig:hcpcfConv}(c), (d) shows convergence for an adaptive refinement strategy. In an adaptive refinement step only a part of the triangles are refined. Since the FEM method works on irregular meshes the refinement of only a fraction of all triangles offers no principal difficulties. In order to choose the triangles which are refined a so called residuum is evaluated on each triangle \cite{ZSC07}. This residuum quantifies the error of the solution on each triangle and only triangles with the largest residuum are refined. According to the definition of the residuum (i.e. the measure for the error) different quantities of the electric field converge with a high rate. Therefore the grid can be refined goal-oriented. E.g. if one is  interested in the field energy one defines a residuum such that this quantity converges with a high convergence rate. This refinement strategy was chosen in Fig. \ref{fig:hcpcfConv}(c), (d). Compared to uniform refinement a certain level of accuracy for the effective refractive index can be achieved with a much smaller number of unknowns. Another example for the quantity of interest used for goal-oriented grid refinement could be the imaginary part of the effective refractive index \cite{ZSC07,POM07a}.

\begin{figure}[t]
(a)\hspace{8.0cm}(b)\\
\includegraphics[width=6.6cm]{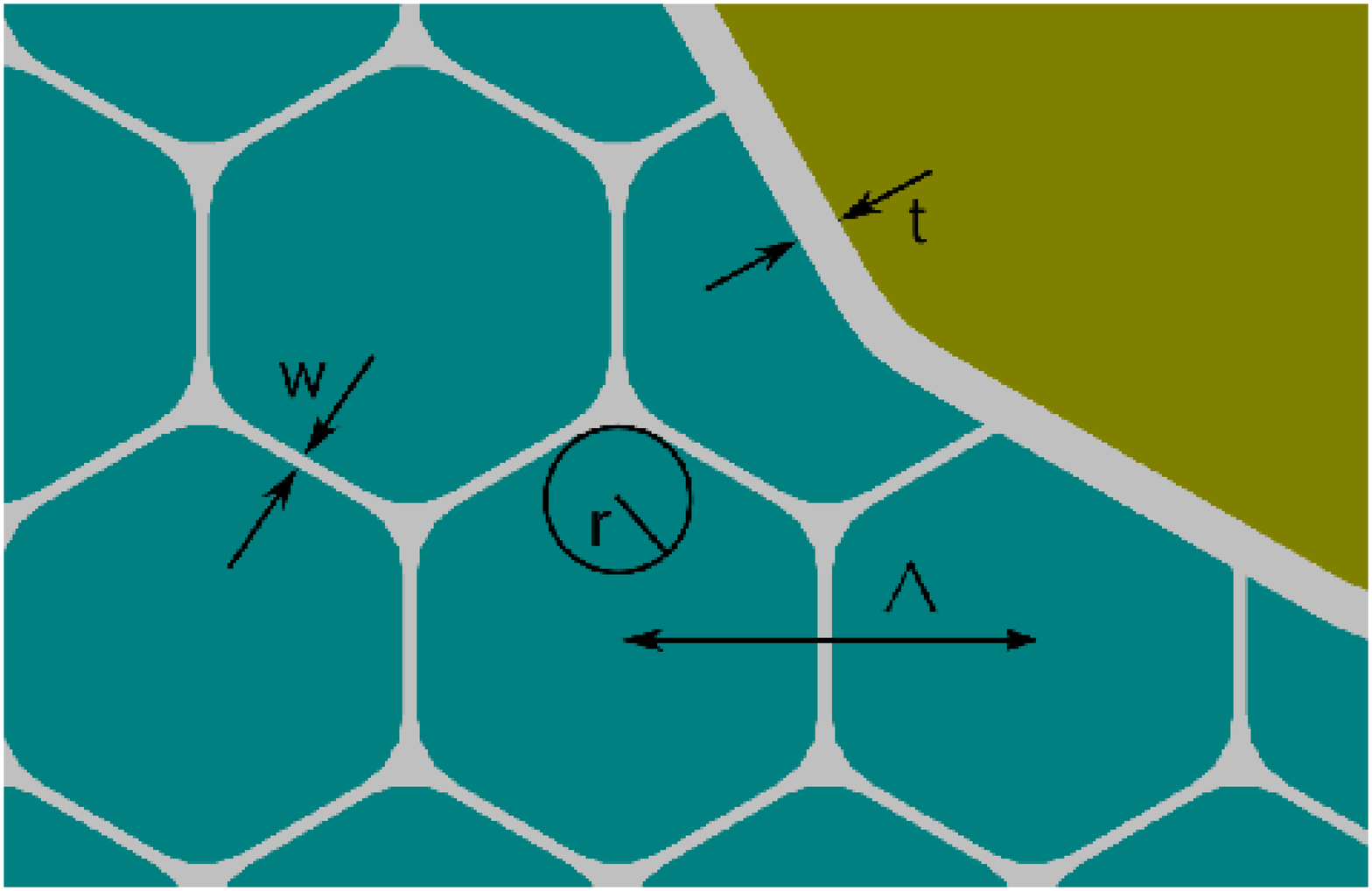}\hfill
\includegraphics[width=6.6cm]{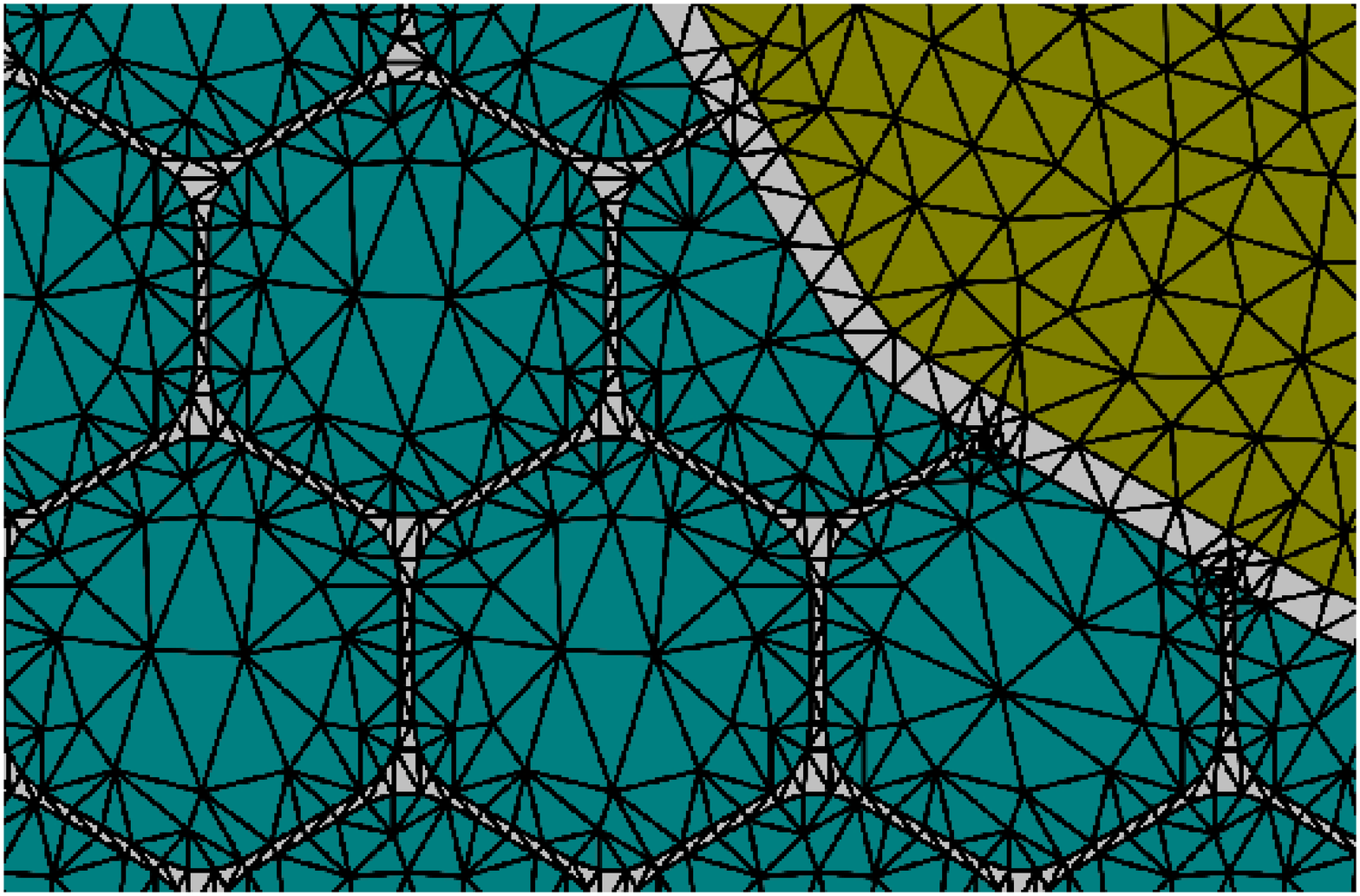}%height=4.5325cm
\caption{\label{fig:hcpcfTriang}(a) geometrical parameters describing HCPCF: pitch $\Lambda$, hole edge radius $r$, strut thickness $w$, core surround thickness $t$; (b) detail from a triangulation of HCPCF. Due to the flexibility of triangulations all geometrical features of the HCPCF are resolved.}
\end{figure}

After we have looked at the convergence we want to optimize the design of a HCPCF in order to minimize radiation losses \cite{POM07}. The fiber we are considering has a hollow core corresponding to 19 omitted hexagonal cladding cells. It is surrounded by hexagonal cladding rings. The cladding rings form a photonic crystal structure which prevents leakage of light from the core to the exterior. With an increasing number of cladding rings the radiation losses therefore decrease, see Fig. \ref{fig:geoScan}(a) \cite{POM07}. We fix the number of cladding rings to 6. The free geometrical parameters are the pitch $\Lambda$, hole edge radius $r$, strut thickness $w$, and core surround thickness $t$ depicted in Fig. \ref{fig:hcpcfTriang}(a) together with the triangulation \ref{fig:hcpcfTriang}(b) \cite{POM07}. Since the finite element method works on irregular meshes the modelling of a complicated structure offers no difficulties.
Fig. \ref{fig:geoScan} shows the radiation losses in dependence on the chosen geometrical parameters keeping all but one fixed in each plot \cite{POM07}. While for the strut thickness $w$ and hole edge radius $r$ we only find one minimum for the pitch $\Lambda$ and core surround thickness $t$ a large number of local minima can be seen. Since with the FEM computation we are able to map the geometrical parameters on the radiation loss we can also use an optimization algorithm to find a geometry with minimal attenuation. Therefore we only have to choose initial values, e.g. from the one dimensional parameter scans Fig. \ref{fig:geoScan}. We fix the hole edge radius to $r=354\,$nm since it has the weakest effect on the radiation losses. For the starting values $\Lambda=1550\,$nm, $t=152\,$nm, $w=50\,$nm optimization yields a minimum value of $\Im(n_{\mathrm{eff}})=5\cdot 10^{-15}\frac{1}{\mathrm{m}}$ for the imaginary part of the effective refractive index. The corresponding geometrical parameters are $\Lambda=1597\,$nm, $w=38\,$nm , $t=151\,$nm \cite{POM07}.

\bibliography{/home/numerik/bzfpompl/myBib}

\begin{thebibliography}{10}

\bibitem{Becker:01a}
R.~Becker and R.~Rannacher.
\newblock An optimal control approach to a posteriori error estimation in
  finite element methods.
\newblock In A.~Iserles, editor, {\em Acta Numerica 2000}, pages 1--102.
  Cambridge University Press.

\bibitem{BerPML}
J.~B\'erenger.
\newblock A perfectly matched layer for the absorption of electromagnetic
  waves.
\newblock {\em J. Comput. Phys.}, 114(2):185--200, 1994.

\bibitem{Burger2006b}
S.~Burger, R.~Klose, A.~Sch\"adle, F.~Schmidt, and L.~Zschiedrich.
\newblock Adaptive {FEM} solver for the computation of electromagnetic
  eigenmodes in 3d photonic crystal structures.
\newblock In A.~M. Anile, G.~Ali, and G.~Mascali, editors, {\em Scientific
  Computing in Electrical Engineering}, pages 169--175. Springer Verlag, 2006.

\bibitem{COU06}
F.~Couny, F.~Benabid, and P.S. Light.
\newblock Large-pitch kagome structured hollow-core photonic crystal fiber.
\newblock {\em Optics Letters}, 31(24):3574--3576, 2006.

\bibitem{CRE99}
R.F. Cregan, B.~J. Mangan, J.C. Knight, P.~St.~J. Russel, P.~J. Roberts, and
  D.C. Allan.
\newblock Single-mode photonic band gap guidance of light in air.
\newblock {\em Science}, 285(5433):1537--1539, 1999.

\bibitem{HOL06}
R.~Holzl{\"o}hner, S.~Burger, P.~J. Roberts, and J.~Pomplun.
\newblock Efficient optimization of hollow-core photonic crystal fiber design
  using the finite-element method.
\newblock {\em Journal of the European Optical Society}, 1(06011), 2006.

\bibitem{LindenEDKZKSBSW06}
S.~Linden, C.~Enkrich, G.~Dolling, M.~W. Klein, J.~Zhou, T.~Koschny, C.~M.
  Soukoulis, S.~Burger, F.~Schmidt, , and M.~Wegener.
\newblock Photonic metamaterials: Magnetism at optical frequencies.
\newblock {\em IEEE Journal of Selected Topics in Quantum Electronics},
  12:1097--1105, 2006.

\bibitem{MON03}
Peter Monk.
\newblock {\em Finite Element Methods for Maxwell's Equations}.
\newblock Oxford University Press, 2003.

\bibitem{NED80}
J.C. Nedelec.
\newblock Mixed finite elements in {R}$^3$.
\newblock {\em Numer. Math.}, 35:315--341, 1980.

\bibitem{POM07}
J.~Pomplun, R.~Holzl{\"o}hner, S.~Burger, L.~Zschiedrich, and F.~Schmidt.
\newblock {FEM} investigation of leaky modes in hollow core photonic crystal
  fibers.
\newblock volume 6480, page 64800M. Proc. SPIE, 2007.

\bibitem{POM07a}
J.~Pomplun, L.~Zschiedrich, R.~Klose, F.~Schmidt, and S.~Burger.
\newblock {F}inite {E}lement simulation of radiation losses in photonic crystal
  fibers.
\newblock submitted to PSS, 2007.

\bibitem{RUS03}
P.~St.~J. Russell.
\newblock Photonic crystal fibers.
\newblock {\em Science}, 299(5605):358--362, 2003.

\bibitem{Schmidt02H}
F.~Schmidt.
\newblock {\em Solution of {I}nterior-{E}xterior {H}elmholtz-{T}ype {P}roblems
  {B}ased on the {P}ole {C}ondition {C}oncept: {T}heory and {A}lgorithms}.
\newblock Habilitation thesis, Free University Berlin, Fachbereich Mathematik
  und Informatik, 2002.

\bibitem{Zschiedrich2005a}
L.~Zschiedrich, S.~Burger, R.~Klose, A.~Sch\"adle, and F.~Schmidt.
\newblock Jcmmode: an adaptive finite element solver for the computation of
  leaky modes.
\newblock In Y.~Sidorin and C.~A. W\"achter, editors, {\em Integrated Optics:
  Devices, Materials, and Technologies IX}, volume 5728, pages 192--202. Proc.
  SPIE, 2005.

\bibitem{ZSC07}
L.~Zschiedrich, S.~Burger, J.~Pomplun, and F.~Schmidt.
\newblock {G}oal {O}riented {A}daptive {F}inite {E}lement {M}ethod for the
  {P}recise {S}imulation of {O}ptical {C}omponents.
\newblock volume 6475, page 64750H. Proc. SPIE, 2007.

\bibitem{Zschiedrich03}
L.~Zschiedrich, R.~Klose, A.~Sch\"adle, and F.~Schmidt.
\newblock A new finite element realization of the {P}erfectly {M}atched {L}ayer
  {M}ethod for {H}elmholtz scattering problems on polygonal domains in 2{D}.
\newblock {\em J. Comput Appl. Math.}, 188:12--32, 2006.

\end{thebibliography}
\bibliographystyle{plain}

\end{document}